\documentclass[traditabstract]{aa}

\usepackage{graphicx}
\usepackage{amssymb}
\usepackage{amsmath}
\usepackage{txfonts}
\usepackage{natbib}
\usepackage{url}
\usepackage{aalongtable}

\newcommand{\teff}{T_\mathrm{eff}}
\newcommand{\alp}{$\alpha$}
\newcommand{\glog}{$\log g$}

\newcommand{\vturb}{$V_{\rm turb}$}
\newcommand{\tauross}{$\tau_{\rm ross}$}
\begin{document}

\title{Photometric signatures of multiple stellar populations in
  Galactic globular clusters} 

\author{L. Sbordone\inst{1,2} \and M.~Salaris\inst{3,1} \and A.~Weiss\inst{1}
        \and S. Cassisi\inst{4}
       }

\institute{Max-Planck-Institut f\"ur Astrophysik,
           Karl-Schwarzschild-Str.~1, 85748 Garching,
           Federal Republic of Germany
           \and
           {GEPI, Observatoire de Paris, CNRS, Universit\'e Paris
             Diderot, Place Jules Janssen, 92190 Meudon, France} 
           \and
           Astrophysics Research Institute, Liverpool John Moores University,
           Twelve Quays House, Egerton Wharf, Birkenhead, CH41 1LD, UK
           \and 
           INAF - Osservatorio Astronomico di Collurania, Via M. Maggini,
           I-64100 Teramo, Italy
           }

\offprints{L. Sbordone (e-mail: lsbordone@mpa-garching.mpg.de)}

\abstract{We have calculated synthetic spectra for typical chemical
  element mixtures (i.e., a standard $\alpha$-enhanced distribution,
  and distributions displaying CN and ONa anticorrelations) found in
  the various subpopulations harboured by individual Galactic globular
  clusters. From the spectra we have determined bolometric corrections
  to the standard Johnson-Cousins and Str\"omgren filters, and finally
  predicted colours. These bolometric corrections and
  colour-transformations, coupled to our theoretical isochrones with
  the appropriate chemical composition, have provided us with a
  complete and self-consistent set of theoretical predictions for the
  effect of abundance variations on the observed cluster
  colour-magnitude diagrams. CNO
  abundance variations affect mainly wavelengths shorter than
  $\sim$400~nm, due to the arise of molecular absorption bands in cooler atmospheres. As a consequence, colour and magnitude changes are
  largest in the blue filters, independently of using broad or
  intermediate bandpasses. Colour-magnitude diagrams involving $uvy$
  and UB filters (and their various possible colour combinations) are thus the ones
  best suited to infer photometrically the presence of multiple
  stellar generations in individual clusters. They are particularly sensitive to 
  variations in the N abundance, with the largest variations affecting the Red Giant Branch (RGB) and lower Main Sequence (MS). 
  BVI diagrams are
  expected to display multiple sequences only if the different
  populations are characterized by variations of the C+N+O sum and/or
  helium abundance, that lead to changes in luminosity and effective
  temperature, but leave the flux distribution above 400~nm
  practically unaffected. A variation of just the helium abundance, up
  to the level we investigate here, 
  affects exclusively the interior structure of stars, and is
  largely irrelevant for the atmospheric structure and the
  resulting flux distribution in the whole wavelength range spanned by
  our analysis.}

\keywords{stars: abundances --Hertzsprung-Russell and C-M diagrams --
  evolution -- globular clusters: general}

\date{Received; accepted}

\authorrunning{Sbordone et al.}
\titlerunning{Multiple stellar populations in globular clusters}
\maketitle

\section{Introduction}

It is now widely accepted that Galactic globular clusters (GCs) host
multiple stellar populations, displaying the effects of cluster
internal chemical evolution.  The majority of GCs appear to have two
approximately coeval subpopulations, or rather, two generations of
stars. The recognition of this fact came with the increasing
observational evidence for what was called \lq{star-to-star abundance
  variations}\rq\ or \lq{abundance anomalies}\rq, which manifest
themselves as variations of
nitrogen, oxygen and sodium, and in some clusters of magnesium and
aluminum abundances. Lithium has been found to vary in a few cases
\citep[\object{NGC~6752}, \object{NGC~6397}, and \object{47
    Tucanae},][]{pasquini05,bonifacio07,lind09,shen10}, anticorrelating with sodium,
and hints exist that elements as heavy as sulfur might be affected
\citep[in \object{47 Tuc}, ][]{sbordone09}.  These abundance
variations are found in stars 
of all evolutionary stages. An extensive discussion of these
anticorrelations, their interpretation as two generations of cluster
stars, and the connection with structural cluster parameters can be
found in \cite{Carretta:10a}, where also a comprehensive list of
original papers is quoted \citep[see also the review
  by][]{gscrev:04}. It is remarkable that the investigation by
\cite{Carretta:10a}, which included over 1200 red giants in 19
carefully selected clusters, revealed that the
\lq{primordial}\rq\ population is generally the less numerous one,
comprising typically 30\% of all cluster members. In contrast,
50--70\% belong to the \lq{intermediate population}\rq, defined as
showing an overabundance of [Na/Fe] of more than $4\sigma$ ($\sigma$
being the typical star-to-star error in [Na/Fe]-determination for each
cluster; see 
\citealt{Carretta:09a}) above the mean abundance observed in field
stars with the same iron content. In some clusters, e.g.\ \object{M13}
and \object{NGC~2808}, the anomalies reach extreme values, in which
case the remaining stars are counted as an \lq{extreme population}\rq,
and -- at least in the case of \object{NGC~2808} (see the following
discussion) -- they could be representative of a third generation of
stars, distinct from the primordial and the intermediate ones.

Such element variations in clusters have been known since a few
decades \citep[e.g., in \object{M3} and \object{M13},][]{Cohen:78}, mainly due to
spectroscopic analyses of cluster members, the number of which has
steadily increased, in particular due to the availability of
high-resolution multi-object spectrographs. Nevertheless, the actual
fraction of cluster stars investigated is still very small. Additional
evidence that GCs host multiple populations came from
high-quality photometric data for thousands of cluster
stars. \citet{Anderson:98} discovered that the broad main sequence
(MS) of the (untypical) cluster \object{$\omega$~Cen} may actually be composed
of two separate sequences.  \citet{lee99} identified such  
sub-populations in the cluster red giant branch (RGB)
and pointed out analogies with \object{NGC 2808} and \object{M22} 
CMDs which were to be confirmed almost a decade later. \citet{bpackmc:04} and follow-up work
demonstrated that the two sequences have different metallicities, with
the bluer one surprisingly hosting the more metal-rich population
($\mathrm{[Fe/H]} = -1.26$ as compared to $-$1.57 of the bulk of the
cluster population). The only explanation found so far
\citep{norris:04,pvbgc:05} is that the bluer sequence has a higher
helium mass fraction $Y\approx 0.38$, compared to a normal value of
0.25.  A weakly populated third main-sequence, redder than the two
others, was found by \citet{villanovaomcen:07} and definitely
confirmed on the basis of an accurate WFC3 photometry by
\citet{bellini:10}.  $\omega$~Cen is also known for hosting at least
four subgiant
\citep{leeomcen:05,sollimaomcen:05,villanovaomcen:07,bellini:10} and
red giant branches \citep[see][for a recent
  compilation]{johnsonomcen:09}, which differ not only in metal
content, but probably also in age.  While these empirical findings,
the cluster mass, and the undisputable presence of a spread in iron
content, definitely put $\omega$~Cen in a special position among GCs, the
photometric evidence for several subpopulations is clear, and has led
to similar investigations in other massive clusters.

As it turned out, multiple cluster sequences in the CMDs are indeed
present in several other clusters. So far, they have been detected in
\object{NGC~2808} \citep[three
  main-sequences;][]{dantona2808:05,piotto2808:07}, \object{NGC~1851} 
\citep[double SGB and RGB][]{milone1851:08,han09}, \object{M4},
\object{NGC~3201} and \object{NGC~1261} \citep[two red giant branch
  populations;][]{marinom4:08, kam10, kravtsov10}, as well as in
\object{NGC~6752} and \object{47 Tuc} \citep[clear spread of the MS,
  also RGB spread in \object{NGC
    6752};][]{anderson:09,milone:10,kravtsov11}.  In a number of cases
\citep[e.g. \object{NGC~1261}, \object{NGC~3201}, \object{NGC~6752},
  see][]{kravtsov10,kam10,kravtsov11}, evidence exists that the two
populations also have different radial distributions, with the redder
RGB population being more centrally concentrated.  Aside from the aforementioned 
\object{NGC 1851}, \citet{piottoiaus258:08}, \citet{marino09} and
\citet{milone:10b} announced preliminary findings of multiple subgiant
branches in seven more clusters, among them \object{M22},
\object{M54}, \object{47 Tuc}, \object{NGC~6388}. 
These object do not display dramatic Fe abundance spread as the one found in \object{$\omega$~Cen}, 
although \object{M22} seems to display an abundance spread of about 0.2 dex, correlating 
with the abundance of s-process elements  \citep{dacosta09,marino09}.
The question therefore is, what the cause for the photometric splitting could be? A
second question concerns the fact that the splitting and/or broadening
of main-sequences, subgiant branches, and red giant branches, is not
present in all photometric colors, a circumstantial evidence that
rules out the presence of a large age difference between the
subpopulations as the cause for the splitting.

The internal chemical evolution of the clusters that led to the
different abundances in the two stellar generations, has affected only
lighter elements up to Al, and can be linked directly to high
temperature proton-capture nucleosynthesis.  In the two scenarios that
potentially could explain the chemical properties, this implies an
increased helium content in the material from which the second
generation is formed. These currently discussed options are winds from
massive stars or mass loss from intermediate mass stars during the
thermally pulsing Asymptotic Giant Branch phase \citep[AGB - see,
  e.g.,][and references therein]{renzini:08}. The increase in helium
could range between being marginal up to the postulated value for the
blue MS in \object{$\omega$~Cen} and the bluest MS in \object{NGC
  2808}, and could be as high as $Y=0.38$
\citep{piotto2808:07}. Therefore both the spectroscopic and the
photometric evidence for multiple cluster populations are intimately
linked to the same mechanism.

For these reasons it is interesting to investigate the photometric
properties of cluster isochrones as they depend on abundance
variations. 
They can be used to separate subpopulations for
further spectroscopic investigations, to verify known or suspected
chemical compositions, or to assess the fraction of first and
second generation stars in a cluster.  
They may also help in understanding the origin of
multiple MS, subgiant- and red giant branches within one cluster.
\citet{swf:06} and \citet{pietrinferni:09} calculated stellar tracks
and isochrones for chemical compositions showing the CN- and
ONa-anticorrelations, as well as for enhanced helium mixtures. They
demonstrated that in the Johnson-Cousins filters V and I only an
extreme helium enhancement leads to a significant colour change as
compared to a standard Pop.~II mixture. They also predicted slight
changes in the RGB bump position, and a separation
of first and second generation stars along the horizontal
branch. \citet{cassisi1851:08} used these isochrones to explain the
splitting of the subgiant branch (SGB) in \object{NGC~1851} as a
consequence of enhanced C+N+O abundances.  \citet{ventura1851:09}
further investigated this question, constraining the total C+N+O
abundance and the maximum helium content compatible with the MS
width. From the observational side, while \citet{yong1851:09} claim a
variation in this sum by a factor of 4, \citet{villanova1851:10}
dispute any evidence for this.  Needless to say that for the
AGB-pollution scenario it is of great importance whether the sum of
the CNO-elements in the second generation stars is constant or has
increased due to He-burning products.

All isochrone computations mentioned so far treated the actual
chemical composition consistently in the stellar interior models. In
particular, appropriate Rosseland mean opacities were
used. \citet{dcfl:07} investigated in detail the influence of
individual element abundances. This limits a strict comparison of
isochrones to the theoretical Hertzsprung-Russell-diagram (HRD).
However, for the precise prediction of colours, also stellar
atmospheres and spectra with consistent chemical composition are
needed. This is of particular importance because there is a clear
evidence for the existence of a correlation between splittings of
evolutionary sequences in colour magnitude diagrams (CMDs) involving
\lq{blue}\rq\ photometric bands (such as the standard U Johnson band)
and abundance anticorrelations, while these photometric peculiarities
disappear when using visual bands like V and I \citep[see among others the case of
  \object{M4},][]{marinom4:08}.

This work presents the first determinations of bolometric corrections
and colours for widely used blue to near infrared standard filters,
from synthetic spectra with chemical compositions typical of both
first and second generation stars found in GCs.  Although we are
postponing a detailed comparison with observed clusters, our results
already allow a basic comparison with observations and specific
predictions.  The structure of the paper is as follows: The next two
sections (2 and 3) present the methodology of our analysis, and a description of
the theoretical model atmosphere and synthetic spectrum calculations,
respectively.  Application of these spectral calculations to
isochrones representative of first and second generation stars
follows in Sect.~4, before a summary and discussion of the results
closes the paper.

\section{Methodology}\label{methods}

\begin{figure}
   \centering
   \includegraphics[width=8.6cm]{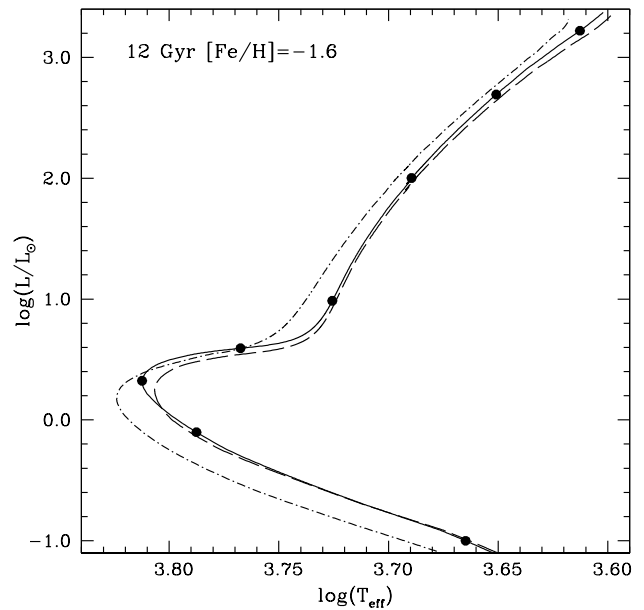}
      \caption{Theoretical isochrones from the MS to the tip of the
        RGB for the four different chemical compositions discussed in
        this paper.  The solid line corresponds to both the reference
        first generation population and the second generation
        population with the CNONa2 metal mixture (both with
        $Y=0.246$), as they are virtually identical. The black dots
        along this isochrone denote the points for which we 
        have calculated model atmosphere and synthetic spectra.  The
        dashed line corresponds to the second 
        generation population with the CNONa1 metal mixture ($Y=0.246$),
        while the dash-dotted line is for the same population, but
        for an initial He mass fraction of $Y=0.40$. See the text for
        more details).}
         \label{HRiso}
   \end{figure}

We have considered a reference isochrone from the BaSTI
database\footnote{The BaSTI database is available at 
\url{http:/www.oa-teramo.inaf.it/BASTI}.} \citep{pietrinferni06},
with the following characteristics: an age of 12 Gyr, helium mass fraction  
$Y=0.246$ and metal mass fraction $Z=0.001$, that results in an iron
abundance of [Fe/H]=$-$1.62 for the $\alpha$-enhanced metal
mixture of the BaSTI models ([$\alpha$/Fe]$\sim$0.4). 
Such an isochrone is representative for
the first generation population in a typical Galactic GC. 
A second generation of GC stars
is represented by coeval (i.e. 12~Gyr old) isochrones calculated for
the same [Fe/H]=$-$1.62, 
$Y=0.246$, but with two different choices for a metal distribution, in
which the elements C, N, O, and Na follow observed
(anti-)correlations. We have also included an additional case for a
mixture with CNONa anticorrelations and, additionally, an increased
initial He abundance of $Y=0.400$ (see Fig.~\ref{HRiso}).

\begin{table*}[ht]
\caption{Mass and number fractions (normalized to unity) for the three
  metal mixtures considered. Full chemical mixtures including all
  elements from H to Ni are listed in Table~\ref{fullmixtures} that
  appears in the online version of the paper. }
\label{zmixtures}     
\centering
{\scriptsize                          
\begin{tabular}{lllllll} 
\hline
   &Reference&          &  CNONa1&          &CNONa2&          \\
   & Number frac.  & Mass frac.  &  Number frac.  & Mass frac  &
Number frac. & Mass frac.  \\ 
\hline
C  & 0.108211 & 0.076451 &  0.013020  & 0.010454 &  0.027200 & 0.019200  \\
N  & 0.028462 & 0.023450 &  0.860012  & 0.805283 &  0.707661 & 0.642054  \\
O  & 0.714945 & 0.672836 &  0.054256  & 0.058031 &  0.113300 & 0.106800  \\
Ne & 0.071502 & 0.084869 &  0.034240  & 0.046189 &  0.071502 & 0.084869  \\
Na & 0.000652 & 0.000882 &  0.001970  & 0.003028 &  0.004110 & 0.005565  \\
Mg & 0.029125 & 0.041639 &  0.013947  & 0.022661 &  0.029125 & 0.041639  \\
Al & 0.000900 & 0.001428 &  0.000431  & 0.000777 &  0.000900 & 0.001428  \\
Si & 0.021591 & 0.035669 &  0.010339  & 0.019412 &  0.021591 & 0.035669  \\
P  & 0.000086 & 0.000157 &  0.000041  & 0.000085 &  0.000086 & 0.000157  \\
S  & 0.010575 & 0.019942 &  0.005064  & 0.010853 &  0.010575 & 0.019942  \\
Cl & 0.000096 & 0.000201 &  0.000046  & 0.000109 &  0.000096 & 0.000201  \\
Ar & 0.001010 & 0.002373 &  0.000484  & 0.001292 &  0.001010 & 0.002373  \\
K  & 0.000040 & 0.000092 &  0.000019  & 0.000050 &  0.000040 & 0.000092  \\
Ca & 0.002210 & 0.005209 &  0.001058  & 0.002836 &  0.002210 & 0.005209  \\
Ti & 0.000137 & 0.000387 &  0.000066  & 0.000210 &  0.000137 & 0.000387  \\
Cr & 0.000145 & 0.000443 &  0.000069  & 0.000241 &  0.000145 & 0.000443  \\
Mn & 0.000075 & 0.000242 &  0.000036  & 0.000132 &  0.000075 & 0.000242  \\
Fe & 0.009642 & 0.031675 &  0.004617  & 0.017238 &  0.009642 & 0.031675  \\
Ni & 0.000595 & 0.002056 &  0.000285  & 0.001118 &  0.000595 & 0.002056  \\
\hline
\end{tabular}}
\end{table*}

The three metal mixtures (mass fraction of metals normalized to unity)
are listed in Table~\ref{zmixtures}.  The $\alpha$-enhanced mixture
employed in the BaSTI database \citep{pietrinferni06}, which
corresponds to typical first generation subpopulations in Galactic
GCs, is labelled as 'reference'. The first mixture representative of
second generation stars is labelled 'CNONa1', and displays -- compared
to the reference $\alpha$-enhanced mixture -- enhancements of N and Na
by 1.8~dex and 0.8~dex by mass, respectively, together with depletions
of C and O by, respectively, 0.6~dex and 0.8~dex.  This is the same
metal distribution already used in the calculations by \citet{swf:06}
and \citet{pietrinferni:09}.  An alternative composition for second
generation stars is labelled 'CNONa2'; it is the same as the CNONa1
mixture but for the enhancement of N that in this case is equal to
1.44~dex by mass.  The important difference between CNONa1 and CNONa2
'second generation' mixtures is that in the first case, at fixed Fe
abundance, the C+N+O mass fraction is enhanced by a factor of 2
compared to the reference composition, whereas the CNONa2 mixture has
the same CNO content (in both number and mass fractions) as the
reference composition, within 0.5\%.  This also implies (given that
the C+N+O sum constitutes most of the total metal content) that for
$Y=0.246$ as in the reference first generation composition, the total
metal mass fraction $Z$ of isochrones for the CNONa1 mixture has to be
larger than the reference composition by a factor 1.84, in order to
have [Fe/H]=$-$1.62.  Both CNONa1 and CNONa2 mixtures are
representative of extreme values of the CNONa anticorrelations
observed in Galactic GCs \citep{carretta05, Carretta:10a}. In
particular, spectroscopic measurements of the C+N+O sum in individual
GCs find typically values in first and second generation stars that
are within a factor of $\sim$2 \citep{carretta05} -- which is the
typical error bar in these estimates.

In case of the CNONa2 mixture, the metal distribution has been
constructed specifically to keep the same $Z$ of the reference
composition, when $Y=0.246$. As for the CNONa1 metal mixture, 
we have also considered the case of an enhanced helium mass fraction
$Y=0.400$. Also in this case we imposed the condition of having
[Fe/H]=$-$1.62 as in the reference composition, which implies a
metallicity $Z$ larger by a factor 1.46. The final mixtures are given
in Table~\ref{fullmixtures}.

\begin{table*}
\caption{Abundances for selected elements in the four chemical
  mixtures used for the atmosphere computations. The full table is made available in the online version.  
\label{fullmixtures}}
     
\centering
{\tiny                         
\begin{tabular}{rll r@{.}l l r@{.}l r@{.}l l r@{.}l r@{.}l lr@{.}l r@{.}l}   
\hline    
   &    & \multicolumn{3}{l}{REFERENCE $Y$=0.246}               &
\multicolumn{5}{l}{CNONa1 $Y$=0.246}&
\multicolumn{5}{l}{CNONa1 $Y$=0.400}               &
\multicolumn{5}{l}{CNONa2 $Y$=0.246} \\ 
\hline
N   & El.& Mass frac. &  \multicolumn{2}{l}{[El]$^{a}$}&   Mass frac. &
\multicolumn{2}{l}{[El]}& \multicolumn{2}{l}{[El]-[El]$_{ref}^{b}$} & Mass frac. &
\multicolumn{2}{l}{[El]}  & \multicolumn{2}{l}{[El]-[El]$_{ref}$} & Mass
frac. & \multicolumn{2}{l}{[El]}  & \multicolumn{2}{l}{[El]-[El]$_{ref}$} \\ 
\hline
 1 & H  & 7.5300e-01 &  12&000 & 7.5217e-01 &  12&000 &  0&0000 & 5.9854e-01 &  12&000 &  0&0000 & 7.5300e-01 &  12&000 &  0&0000 \\ 
 2 & He & 2.4600e-01 &  10&915 & 2.4600e-01 &  10&916 &  0&0005 & 4.0000e-01 &  11&226 &  0&3108 & 2.4600e-01 &  10&915 &  0&0000 \\ 
...&    & \multicolumn{12}{l}{  }                                                                                                 \\ 
 6 & C  & 7.6386e-05 &   6&930 & 1.9172e-05 &   6&330 & -0&5999 & 1.5256e-05 &   6&330 & -0&5999 & 1.9184e-05 &   6&330 & -0&6001 \\ 
 7 & N  & 2.3430e-05 &   6&350 & 1.4768e-03 &   8&150 &  1&8000 & 1.1752e-03 &   8&150 &  1&8000 & 6.4150e-04 &   7&787 &  1&4374 \\ 
 8 & O  & 6.7226e-04 &   7&750 & 1.0642e-04 &   6&950 & -0&8000 & 8.4687e-05 &   6&950 & -0&8000 & 1.0671e-04 &   6&951 & -0&7993 \\ 
...&    & \multicolumn{12}{l}{  }                                                                                                 \\  
11 & Na & 8.8125e-07 &   4&710 & 5.5531e-06 &   5&510 &  0&7999 & 4.4189e-06 &   5&510 &  0&7999 & 5.5602e-06 &   5&510 &  0&8000 \\ 
12 & Mg & 4.1603e-05 &   6&360 & 4.1558e-05 &   6&360 &  0&0000 & 3.3070e-05 &   6&360 &  0&0000 & 4.1603e-05 &   6&360 &  0&0000 \\ 
13 & Al & 1.4268e-06 &   4&850 & 1.4249e-06 &   4&850 & -0&0001 & 1.1339e-06 &   4&850 & -0&0001 & 1.4268e-06 &   4&850 &  0&0000 \\ 
14 & Si & 3.5638e-05 &   6&230 & 3.5600e-05 &   6&230 &  0&0000 & 2.8329e-05 &   6&230 &  0&0000 & 3.5638e-05 &   6&230 &  0&0000 \\ 
...&    & \multicolumn{12}{l}{  }                                                                                                 \\  
20 & Ca & 5.2045e-06 &   5&240 & 5.2009e-06 &   5&240 &  0&0002 & 4.1387e-06 &   5&240 &  0&0002 & 5.2045e-06 &   5&240 &  0&0000 \\ 
...&    & \multicolumn{12}{l}{  }                                                                                                 \\ 
22 & Ti & 3.8667e-07 &   4&034 & 3.8512e-07 &   4&033 & -0&0013 & 3.0646e-07 &   4&033 & -0&0013 & 3.8667e-07 &   4&034 &  0&0000 \\ 
...&    & \multicolumn{12}{l}{  }                                                                                                 \\ 
26 & Fe & 3.1648e-05 &   5&880 & 3.1613e-05 &   5&880 &  0&0000 & 2.5156e-05 &   5&880 &  0&0000 & 3.1648e-05 &   5&880 &  0&0000 \\ 
...&    & \multicolumn{12}{l}{  }                                                                                                 \\ 
28 & Ni & 2.0542e-06 &   4&671 & 2.0503e-06 &   4&670 & -0&0003 & 1.6315e-06 &   4&670 & -0&0003 & 2.0542e-06 &   4&671 &  0&0000 \\ 
\hline    
   & $Z$   & \multicolumn{3}{l}{{\bf 9.9937e-04}}      &
\multicolumn{5}{l}{{\bf 1.8343e-03}}&               \multicolumn{5}{l}{
 {\bf 1.4597e-03}}               &   \multicolumn{5}{l}{{\bf 9.9937e-04}} \\ 
\hline
\multicolumn{14}{l}{\scriptsize $a$ [El]=$\log N(\mathrm{El})-\log N(\mathrm{H})+12$}\\
\multicolumn{14}{l}{\scriptsize $b$ [El] for this mixture minus [El] for the reference mixture.}\\
\end{tabular}}
\end{table*}

Strictly speaking, our choice of keeping [Fe/H]=$-$1.62 for the CNONa1
mixture with enhanced helium is not in agreement with scenarios that
could explain the chemical signatures of the second generation
stars. In these models Fe and all other iron-peak elements are not
affected, and indeed the absolute iron-abundance should be kept
constant, so that an increase of $Y$, accompanied by a decrease in
hydrogen abundance, therefore would affect [Fe/H].  So far, there is
only one highly 
significant empirical evidence showing the existence of differences in
[Fe/H] between distinct subpopulations within a GC, e.g, the case of
\object{NGC~2808} \citep{bragaglia10}.  This has been interpreted as due to
significant changes of the initial helium abundance between the
various subpopulations, already discussed in the introduction.  It is
also important to notice that almost all chemical abundance studies so
far were based on scaled solar composition model atmospheres, at most
including $\alpha$-enhancement. If a population displays a large
enough departure from the scaled solar composition, a failure to
account for the true chemical pattern in the model atmosphere could
potentially skew the derived [Fe/H], possibly masking differences in
[Fe/H] among populations.

At any rate, even for our choice of an extreme He enhancement, by varying $Y$ keeping $Z$ fixed one would obtain a
variation of [Fe/H] of about 0.1~dex. We have verified from the
$\alpha$-enhanced \citet{castelli04} models, that such a small
variation of [Fe/H] affects the bolometric corrections to the
photometric filters discussed in this work at the level of at most
$\sim$0.01-0.02~mag, and only at blue wavelengths.

The HRD from the MS to the tip of the RGB of the isochrones considered
in our analysis is shown in Fig.~\ref{HRiso}.  In case of the CNONa2
mixture, the corresponding isochrone is practically identical to the reference
$\alpha$-enhanced one, therefore we have used this latter isochrone to 
represent also a second generation population born with the CNONa2 mixture. 
As we have tested with specific
calculations -- using the same input physics as in \citet{swf:06} --
as long as the C+N+O sum is unchanged (in the CNONa2 mixture it is within 
0.5\% of the reference metal distribution), an isochrone calculated using a
metal mixture with typical GC CNONa anticorrelations is equal to a
standard $\alpha$-enhanced isochrone with the same [Fe/H], which in 
case of the  CNONa2 mixture corresponds to having the same $Z$. 
For the two CNONa1 mixtures with initial He abundances of $Y=0.248$ and
$Y=0.400$ we used isochrones also taken from the BaSTI
database \citep{pietrinferni:09}, to represent subpopulations with
extreme values of the CNONa  anticorrelations and a large He
enhancement.  The CNONa1, $Y=0.246$ isochrone is essentially
identical to the reference one but for the TO and subgiant branch
(SGB) regions, as discussed in \citet{swf:06} and
\citet{pietrinferni:09}. The TO is fainter and redder, and the SGB is
fainter compared to the reference isochrone. When helium is enhanced,
as is well known, the MS, TO and -- to a lesser degree -- the RGB become
hotter. 

Along the reference isochrone we have
selected 8 key points (marked as black dots in Fig.~\ref{HRiso})
that cover almost the full range of $\teff$ and luminosities. For each
of these points we have calculated appropriate model atmospheres and
synthetic spectra for each of the four metal mixture/He mass fraction
pairs described before.  The parameters of the model atmosphere
calculations are reported in Table~\ref{parameters}. The calculations
denoted as \lq{main set}\rq\ have been employed to produce the theoretical
CMDs for first and second generation stars displayed in
Sect.~\ref{comp}, while the `test' calculations have been used to
determine the effect of the choice of microturbulence on the
results.

The next step of our analysis has been to produce observational CMDs
in several photometric bands, starting from the isochrones displayed
in Fig.~\ref{HRiso}. We have considered the Johnson-Cousins UBVI and
the Str\"omgren $uvby$ photometric systems, that cover the full
wavelength range spanned by our spectra, and allow us to study the
effect of the different metal mixtures on both broad- and
intermediate-band filters.

For this purpose we have first produced CMDs in selected filter
combinations for all
isochrones, by employing the \citet{castelli04} bolometric correction
tables, for the reference $\alpha$-enhanced composition \citep[see
  also][]{cassisi04}.  From our own synthetic spectra calculations,
for each of the 8 key points we have then calculated the BC differences
($\Delta$BC) between the reference metal mixture and each of the other
mixtures representing second generation stars. This enables us to
study the effect of the abundance correlations on the bolometric
corrections at fixed $\teff$ and surface gravity.  We have then
applied these corrections $\Delta$BC -- by interpolation in $\teff$
between our key points -- to the isochrones with the chemical
compositions of second generation stars.

One important issue to notice is that the $\Delta$BC values determined
for the chosen key points are strictly speaking not applicable to the
two isochrones for the CNONa1 mixture in Fig.~\ref{HRiso}.  In case of
a CNONa1 metal mixture and $Y$=0.246, along the turn off and SGB
regions, a given value of $\teff$ corresponds to a different surface
gravity (because of a different luminosity and slighlty different
evolving mass) than the reference (and the CNONa2)
isochrones. However, as we have tested with some sample calculations,
these differences in gravities, that amount to 0.08~dex at most, do
not affect appreciably the values of $\Delta$BC at a fixed $\teff$.

Second, all along the CNONa1 isochrone with $Y=0.40$, the surface
gravity differs by $\sim$0.2-0.4~dex -- at fixed $\teff$ -- from our
key points, because of different evolving mass and luminosities. We
have verified with some sample calculations, that changes in surface
gravity of this order affect the corrections $\Delta$BC by at most
$\sim$0.01~mag, and only for the $u$ and U filters. Furthermore, the
TO region of this He-enhanced isochrone is hotter than our hottest key
point.  We have therefore applied a linear extrapolation (by at most
150~K) to our grid of $\Delta$BC values to cover the relevant
temperature range.  We have verified that we have not introduced any
appreciable systematic error, by comparing the extrapolated $\Delta$BC
values with values obtained from appropriate spectra calculated
for the TO point of the $Y=0.40$, CNONa1 isochrone. 

Finally, we point out that a similar linear extrapolation of $\Delta$BC (by at
most 100~K) has been applied to cover the last $\sim$0.10-0.15~dex in
bolometric luminosity, close to the RGB tip, a region that is only
very sparsely populated in the CMDs of typical Galactic GCs. 
To summarize, we
are sure that the application of our $\Delta$BC values to the CNONa1
mixture isochrones is not introducing any significant error, which
would affect our conclusions.

   \begin{figure*}
   \centering
   \includegraphics[width=17cm]{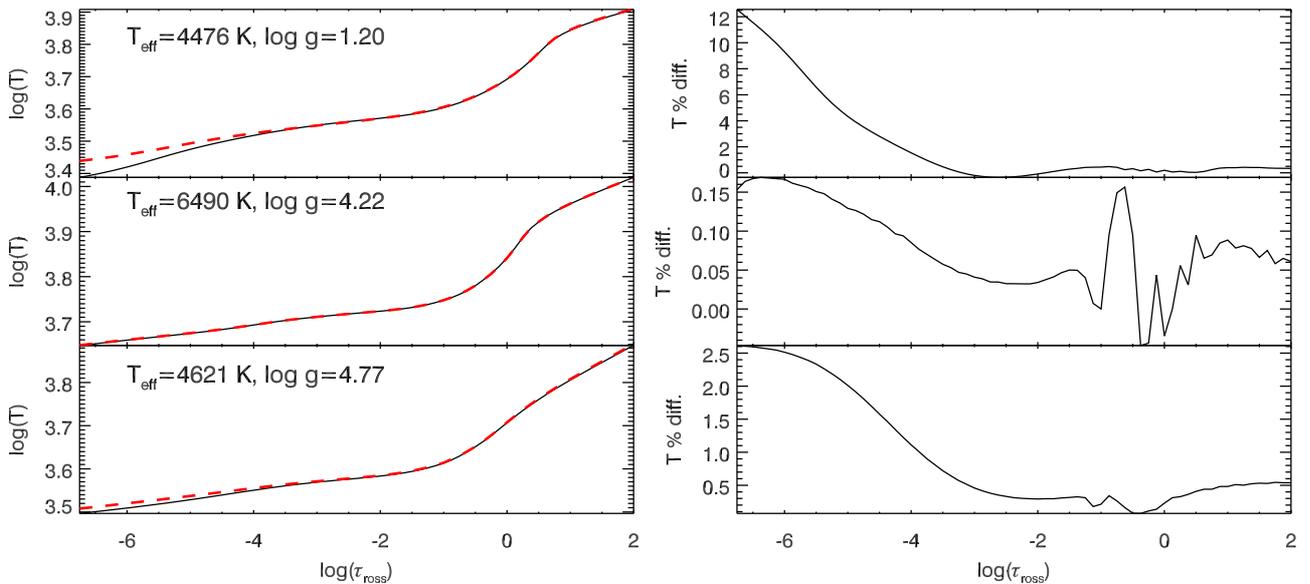}
      \caption{The left panel displays the temperature structures for
        a MS, TO, and a RGB atmosphere as used in the
        present work.  The black continuous line represents the model
        computed with the reference chemical mixture, the red dashed
        line the one assuming the CNONa1 mixture. In both cases,
        $Y=0.246$. The right panel shows the
        percentage temperature difference between the two different
        chemical compositions.} 
         \label{tstruct}
   \end{figure*}


\section{Atmosphere models and synthetic fluxes.}

The synthetic spectra for the selected key points have been derived
from self-consistent model atmospheres and synthetic spectra computed
with the ATLAS~12 and SYNTHE
\citep{kurucz05,castelli05,sbordone05,sbordone07} codes, respectively.
ATLAS~12 employs opacity sampling to estimate the line opacity in the
atmosphere, allowing the computation of model atmospheres for arbitrary
chemical compositions. For each considered chemical mixture and each
set of atmospheric parameters ($\teff$, \glog, \vturb) an ATLAS~12
plane-parallel, LTE model atmosphere was computed, and a SYNTHE
spectral synthesis was performed between 300 nm and 1000 nm. The
spectral synthesis accounts for the full set of atomic and molecular
lines included in the \citet{sbordone05} and \citet{sbordone07} Linux
ATLAS port\footnote{We {\em did not} include TiO lines in the
  calculations, after verifying that no significant TiO bands formed
  in some representative cases, given that they would have made the
  calculations remarkably heavier.}, as well as the predicted lines
normally used for pre-tabulated opacities calculation
\citep{kurucz05_2}, for the purpose of this spectral synthesis was to
derive intermediate-to-broad band colours. {\em All the computed
  models and synthetic spectra will be made available online to the
  community\footnote{For this preprint, we have not yet set up a download 
  system which will be made available for the final publication. 
  In the meantime, anyone interested might contact us directly.}.}

\subsection{Producing the full chemical mixtures}

The first step of our calculations was to provide the full chemical
mixtures for the atmospheres. The three chemical compositions of
Table~\ref{zmixtures} were derived from the stellar interior
calculations, and included only those elements relevant for that
purpose in the form of mass fractions of metals normalized to
unity. We will henceforth use the term \lq{interior}\rq\ when speaking
about this restricted set of elements.

While it is possible to consider only a subset of elements for the
stellar interior modeling without adversely influencing the results,
abundances for all remaining metals should be included in synthetic
spectrum calculations. The fractions of elements in the interior
mixture should then be converted into absolute abundances, including
appropriate values for H and He, and the missing elements added. This
process is essentially arbitrary, since there is no single way to fill
in the missing metal abundances. We proceeded as follows:
\begin{itemize}
\item The interior metal mixture, as listed in Table~\ref{zmixtures},
  was converted to a full metal mixture by filling in all the elements
  except H and He.  To do so, we took as reference the chemical
  mixture of an \alp-enhanced, [Fe/H]=-1.5 model from the
  \citet{castelli04} grid\footnote{Available online at
    {\url{http://wwwuser.oat.ts.astro.it/castelli/grids.html}}}.  The
  total mass fraction of the elements considered for the interiors
  amounts to 0.986 in this mixture. Thus, their abundances in the
  interior mixtures of Table~\ref{zmixtures} were scaled by this
  factor, and the remaning 1.4\% distributed over the missing elements
  according to their abundance fractions in the Castelli mixture.
  This has produced the three full metal mixtures of Table
  \ref{fullmixtures}.
\item The helium mass fraction $Y$ was set to the two possible values,
  $Y=0.246$ or $Y=0.400$. To produce a full set of abundances we then
  only needed to set a ratio between the H mass fraction $X$ and the
  cumulative mass fraction of metals, $Z$.  This was done by {\em
    enforcing} [Fe/H]=-1.62.  We came to such choice since
  observational evidence is that stars from different populations in
  GCs usually share the same [Fe/H] (see also the discussion in
  Sect.~2).  The final full mixtures used for the atmospheres are
  listed in Table~\ref{fullmixtures}.
\end{itemize}

\subsection{The model atmospheres}
\label{themodelatmospheres}

ATLAS~12 has been employed to produce sets of model atmospheres from
lower MS to bright RGB stars -- the parameters are listed in
Table~\ref{parameters} -- for each of the four chemical compositions
given in Table~\ref{fullmixtures}.

\begin{table}
\caption{Parameters of the model atmospheres. Calculations with the
  'main set' of parameters have been employed to produce theoretical
  CMDs for first and second generation stars. Calculations with the
  'test' parameter sets have been used to estimate the effect of
  microturbulence.}         
\label{parameters}     
\centering                          
\begin{tabular}{cccc}       
\hline                
$T_\mathrm{eff}$ & $\log g$ & $V_\mathrm{turb}$ &  Mixtures \\    
K & $\mathrm{cm/s^2}$   & km/s &                \\
\hline 
main set & & & \\
\hline                      
4100  & 0.50  & 2.0 & all \\
4476  & 1.20  & 2.0 & all \\
4892  & 2.06  & 2.0 & all \\
5312  & 3.21  & 2.0 & all \\
5854  & 3.78  & 2.0 & all \\
6490  & 4.22  & 2.0 & all \\
6131  & 4.50  & 2.0 & all \\
4621  & 4.77  & 2.0 & all \\
\hline
 $V_\mathrm{turb}$ test  & & & \\
\hline
4476  & 1.20 & 0.5 & { ref. Y=0.246, CNONa2 Y=0.4} \\
4476  & 1.20 & 1.0 & { ref. Y=0.246, CNONa2 Y=0.4} \\
4476  & 1.20 & 1.5 & { ref. Y=0.246, CNONa2 Y=0.4} \\
6490  & 4.22 & 0.5 & { ref. Y=0.246, CNONa2 Y=0.4} \\
6490  & 4.22 & 1.0 & { ref. Y=0.246, CNONa2 Y=0.4} \\
6490  & 4.22 & 1.5 & { ref. Y=0.246, CNONa2 Y=0.4} \\
4621  & 4.77 & 0.5 & { ref. Y=0.246, CNONa2 Y=0.4} \\
4621  & 4.77 & 1.0 & { ref. Y=0.246, CNONa2 Y=0.4} \\
4621  & 4.77 & 1.5 & { ref. Y=0.246, CNONa2 Y=0.4} \\
\hline   
\end{tabular}
\end{table}

Figure~\ref{tstruct} compares the temperature stratification in models
employing the reference and CNONa1 ($Y=0.246$) mixtures. Shown are
temperature (left panel) and the corresponding relative differences
(right panel) against Rosseland optical depth for three 
models corresponding to a cool MS, a TO and an upper RGB
star. Significant differences arise only in the outermost layers and
only in cool models.  As we will show below, these differences arise
from the generally lower molecular absorption in the reference mixture
compared to the other models, which affects mostly the outer layers of
the cooler atmospheres. The strongest differences arise in the cool
giants and are still appreciable in the coolest dwarf models, but
become negligible in the warmer stars close to the TO or along the
SGB. In the $\teff=4476$~K giant, differences range from about 50~K at
\tauross=$10^{-4}$ to about 250~K at \tauross=$10^{-6}$, which is the
highest value encountered amongst the models explored in this work.

   \begin{figure}
   \centering
   \includegraphics[width=8cm]{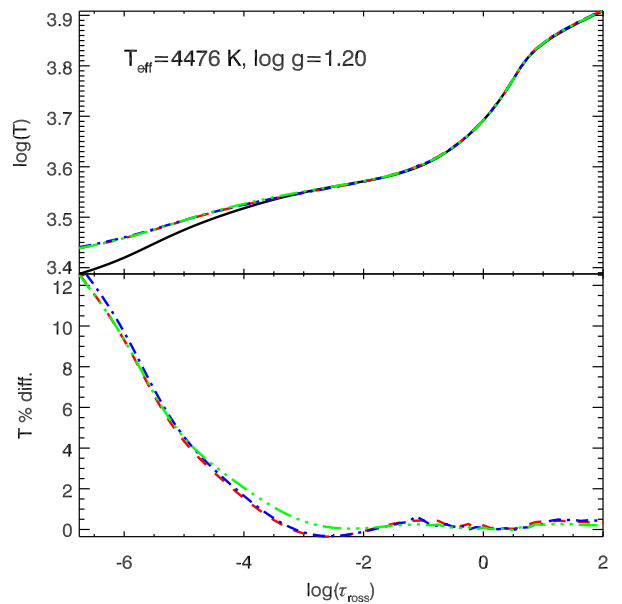}
      \caption{The upper panel compares the temperature stratification
        for the four $\teff=4476$~K, \glog=1.2 giant models.  The
        solid line denotes the reference mixture, the dashed
        line the CNONa mixture with $Y=0.246$, the dot-dashed line the CNONa1
        mixture with $Y=0.4$, and the dash-dot-dot line the CNONa2
        mixture. In the
        lower panel the percentage temperature difference for
        the last three mixtures with respect to the reference one is 
        plotted; colour and line type are coded as in the upper panel.}
         \label{fourcases}
   \end{figure}

   \begin{figure*}
   \centering
   \includegraphics[width=17cm]{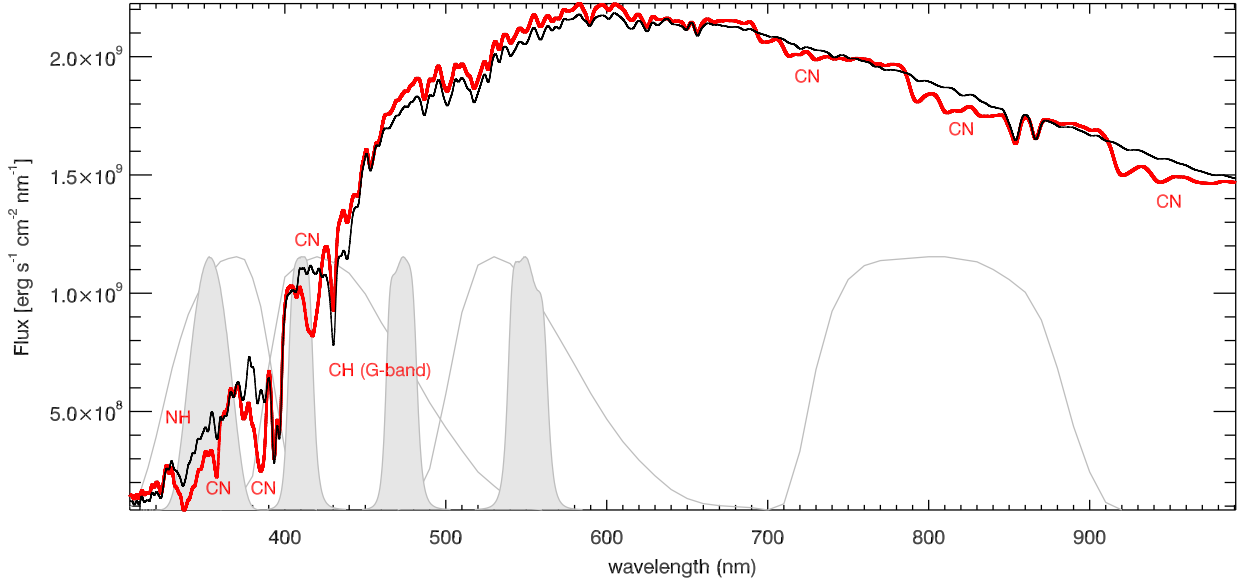}
      \caption{The flux distribution for the RGB model with
        $\teff=4476$~K and \glog=1.2 for the reference mixture (black)
        and the CNONa1 mixture (red), both with the standard helium
        content of $Y=0.246$. We have superimposed the transmission
        curves for the Johnson-Cousins U, B, V, and I filters (thin,
        black lines; left to right) and for the Str\"ogren $uvby$
        filtes (grey-shaded regions).  The synthetic spectra are
        broadened by convolution with a gaussian of FWHM=1700~km/s for
        readability purposes, and in order to roughly match the
        resolution of the \citet{castelli04} ATLAS9-based fluxes. A
        number of molecular bands which vary significantly between the
        two mixtures are labeled by the name of the corresponding
        molecule.}
         \label{RGBflux}
   \end{figure*}

Figure~\ref{fourcases} displays the temperature stratification of the
RGB models shown in Fig. \ref{tstruct} for all our four mixtures, as
well as their deviations from the case of the reference mixture. The
results are representative of the general result for all models. Even
large variations in the He abundance have a quite minor effect on the
atmosphere structure (as is evident from the extremely similar
structures in the CNONa1 $Y=0.246$ and $Y=0.4$ cases). It might be worth
reminding that this does not mean that the helium content will not affect the
colours of second generation stars, since varying the He content has
an important effect on the stellar interiors (Fig.~\ref{HRiso}). 
It merely indicates that the helium
content does not affect these cool atmospheres, and thus the
relationship between stellar parameters and colours. Passing from the
CNONa1 to the CNONa2 mixture has a slightly more noticeable effect,
but we want to stress that the difference never exceeds 25~K.

   \begin{figure*}
   \centering
   \includegraphics[width=17cm]{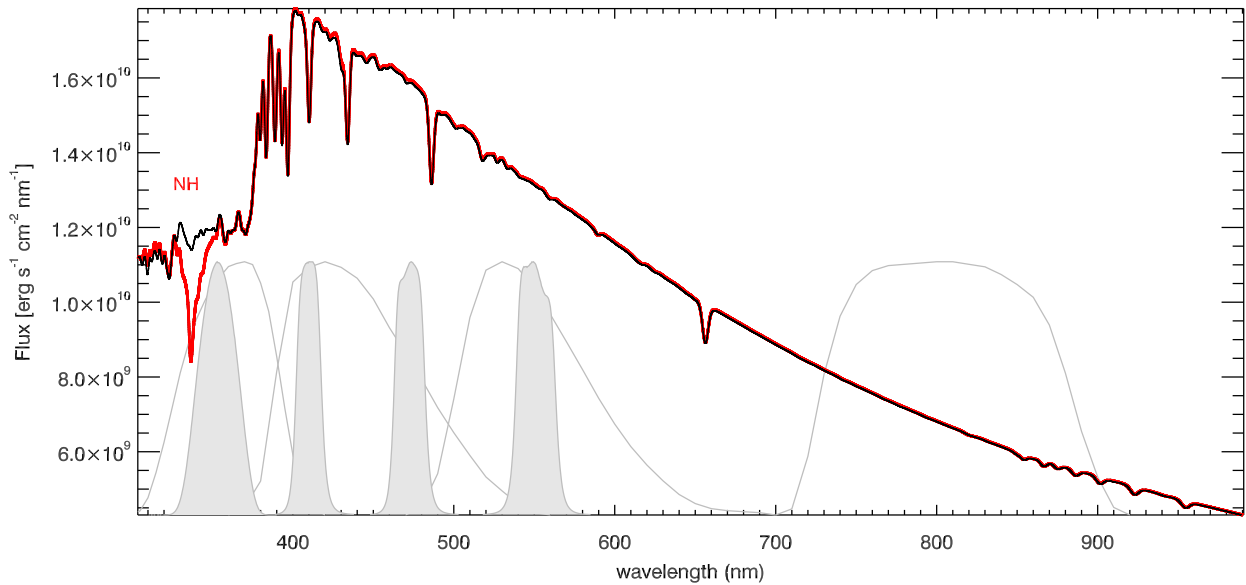}
      \caption{As in Fig.~\ref{RGBflux} but now for the TO model with
        $\teff=6490$~K and \glog=4.22.} 
         \label{TOflux}
   \end{figure*}

As far as the model atmospheres are concerned, one can thus conclude
that the explored chemical mixtures do not have a major effect on the
atmospheric structure, except in the outermost layers, where -- at any
rate -- LTE models such as the ones employed here are less physically
sound.

\subsection{The synthetic spectra}

\begin{figure*}
   \centering
   \includegraphics[width=17cm]{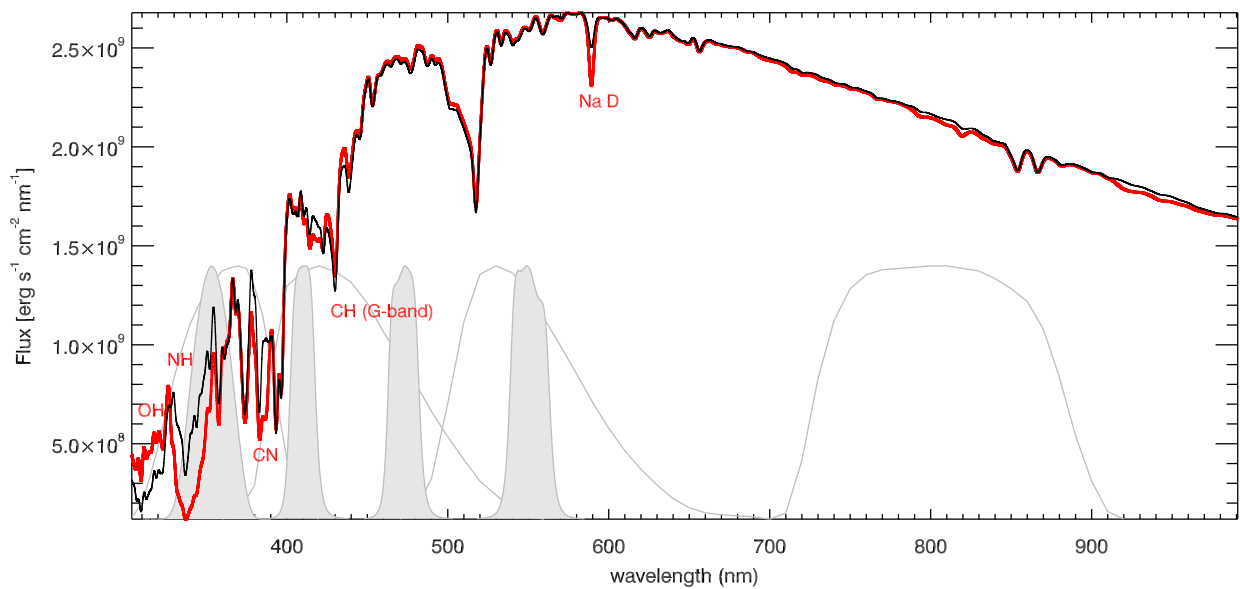}
      \caption{As in Fig.~\ref{RGBflux} but now for the TO model with
        $\teff=4621$~K and \glog=4.47. Here, the Na D doublet is also
        labeled}
         \label{MSflux}
   \end{figure*}

We display in Figs.~\ref{RGBflux}, \ref{TOflux}, and \ref{MSflux} the
synthetic fluxes computed for the same three sets of parameters as in
Fig.~\ref{tstruct}, as representative of the effects of changing
chemical composition.  More specifically, we display the case of the
reference (black) and CNONa1 (red) mixtures, both with
$Y=0.246$. Synthetic spectra were computed with very fine sampling
($\Delta \lambda / \lambda = 300000$) but are shown here after
convolution with a FWHM=1700~km/s Gaussian, for readibility purposes,
and in order to roughly match the resolution of \citet{castelli04}
fluxes.  In each plot we label some prominent absorption features
(mostly molecular bands), whose strength changes significantly among
chemical mixtures.

Figure~\ref{RGBflux} compares the synthetic spectra of a
$\teff=4476$~K, \glog=1.2 giant. The CNONa1 mixture shows much
stronger NH and CN absorption bands in the U, B, and I filters. This
is essentially due to the much higher N abundance of the CNONa1
mixture, despite the fact that the C abundance is lower. This indicates
that the N abundance acts as bottleneck in forming CN
molecules. Conversely, the G-band, a CH feature falling into the B
filter, appears stronger in the reference mixture, since the C
abundance is higher. The increased opacity in the blue part of the
spectrum for the CNONa1 mixture leads to the increase of the continuum
flux redwards of about 450 nm, which explains the overall higher flux
observed in the CNONa1 spectrum between 450 and 690 nm, and further in
the red in the intervals along the strong CN absorption bands.

In the spectrum of the TO atmosphere ($\teff=6490$~K, \glog=4.22,
Fig.~\ref{TOflux}) the much higher temperature prevents the formation
of CN and CH alike, thus removing the most prominent causes for
increased blue opacity in the CNONa1 case, as well as most of the
influence of the G-band. Only the very strong NH band around 340~nm is
still visible. The much more similar
spectra for the two mixtures in the TO atmosphere are reflected in the
extremely similar atmospheric structures displayed by the two TO
models in Fig.~\ref{tstruct}.

The cool MS star ($\teff=4621$~K, \glog=4.77, Fig. \ref{MSflux})
displays features similar to the RGB one, but shows also a much
stronger OH absorption at the blue edge of the U filter range in the
reference case, due to the higher O abundance: the same band is
visible, but much less prominent, also in the MS and RGB stars. The Na
D doublet becomes strongly wing-dominated in this spectrum and thus
appears much stronger in the CNONa1 spectrum, where the Na abundance
has been increased to reproduce the observed Na-O anticorrelation.
Red CN bands have an
almost negligible effect in this star. The strong absorption feature
around 415~nm is the MgH A-X band.

\begin{figure}
   \centering
   \includegraphics[width=9cm]{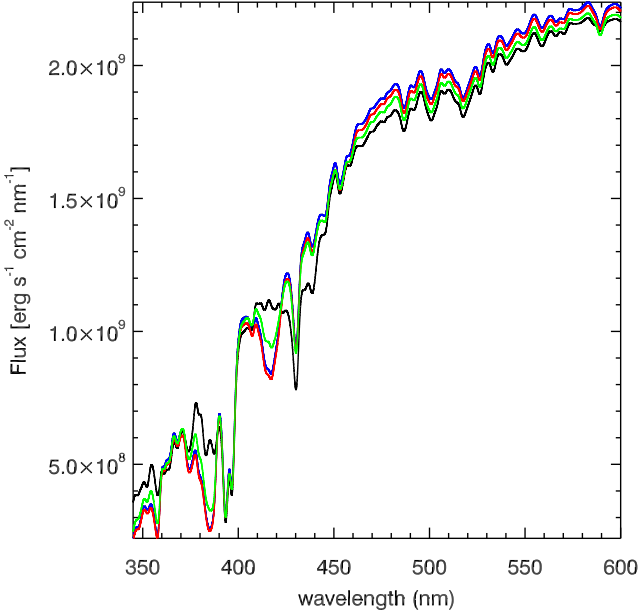}
      \caption{The range between 300 and 600~nm is displayed for all
        mixtures considered, in the case of the $\teff=4476$~K, \glog=1.2
        giant atmosphere. Black line: reference mixture; red: CNONa1 ($Y=0.246$); 
        blue: CNONa1 ($Y=0.400$); green:
        CNONa2 ($Y=0.246$).} 
         \label{fourmixtures}
   \end{figure}

In Fig.~\ref{fourmixtures} we finally show how the blue-visible part
of the spectrum changes among all the four mixtures considered, in the
same RGB stellar model as described above. Two effects are immediately
apparent: first, the CNONa2 mixture -- sharing the same $Z$ as the
reference one, as well as a lower N enhancement -- shows less prominent
NH and CN bands than the CNONa1 mixture, as well as a continuum which
resembles more the one of the reference mixture. Second, the change in
the He abundance between the two CNONa1 mixture has only a tiny effect
on the flux distribution. It has here to be considered that our choice
of conserving [Fe/H] among mixtures leads to a {\em lower} Fe
abundance (and $Z$, for that matter) in the CNONa1 mixture with $Y=0.400$
with respect to the one with $Y=0.246$, thus mitigating the main effect
of increasing He, i.e.\ an increase in molecular weight.

Our synthetic fluxes have been compared with MARCS synthetic fluxes \citep{gustafsson08},
obtaining equivalent results for the reference composition. MARCS ``CN-cycled'' fluxes 
are produced with CNO abundances that depart from the standard mixture less than the ones we considered here, 
which, as expected, leads to a behavior which is intermediate between our reference and CNONa1/2 cases.

\section{Bolometric corrections and colour-magnitude-diagrams}\label{comp}

\begin{figure}
\centering
\includegraphics[width=8.6cm]{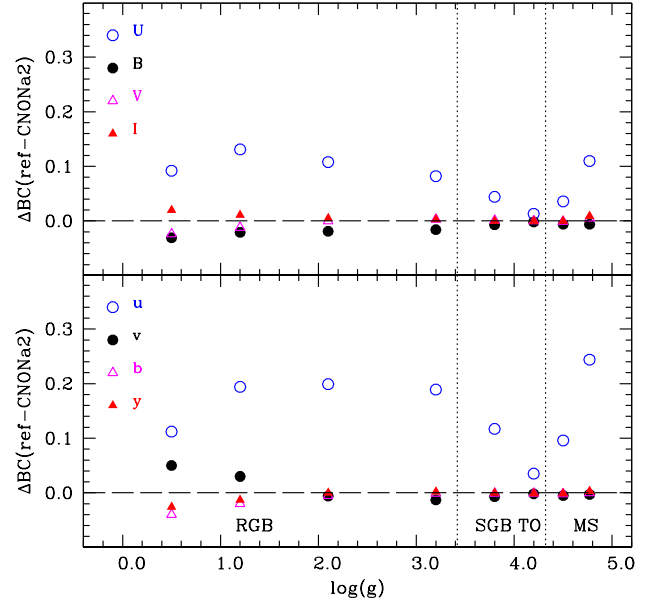}
\caption{{\sl Upper panel}: Difference between the UBVI bolometric
  corrections $\Delta$BC for the reference mixture and the CNONa2
  mixture with the same $Y=0.246$.  {\sl Bottom panel}: As in the upper
  panel, but for the $uvby$ system.}  
\label{BC1}
\end{figure}

\begin{figure}
\centering
\includegraphics[width=8.6cm]{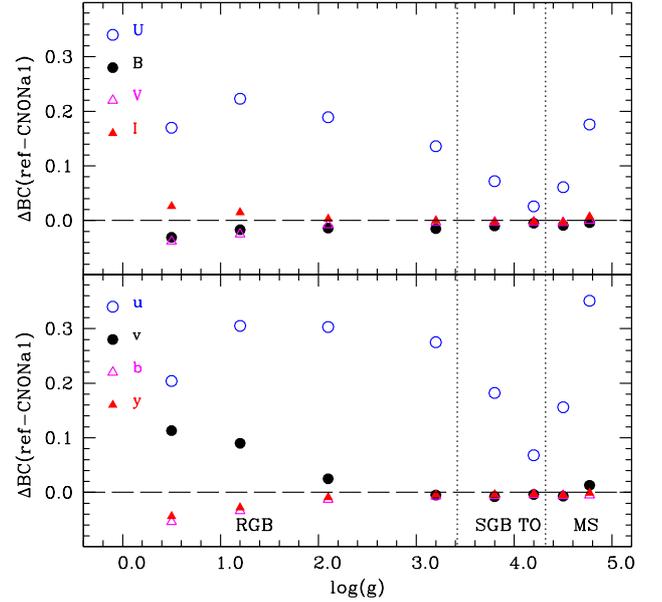}
\caption{As in Fig.~\ref{BC1} but for the CNONa1 mixture.}
\label{BC2}
\end{figure}

\begin{figure}
\centering
\includegraphics[width=8.6cm]{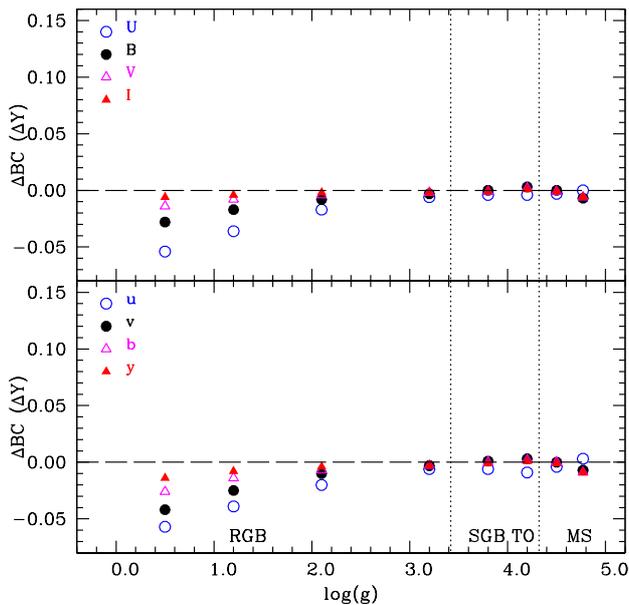}
\caption{As in Fig.~\ref{BC1}, but for $\Delta$BC calculated between
  the CNONa1 mixture with $Y=0.246$, and the same metal mixture with
  $Y=0.400$.} 
\label{BC3}
\end{figure}

As discussed in Section~\ref{methods}, we have calculated the BC
differences $\Delta$BC for each of the key points, in the UBVI, and
$uvby$ photometric filters.  We followed the method presented in
\citet{girardi02} using passband definitions from \citet{bessell90}
and \citet{stroem56}.  The results for the different chemical
compositions are displayed in Figs.~\ref{BC1}-\ref{BC3}.
Figure~\ref{BC1} shows $\Delta$BC for the CNONa2 mixture as a function
of \glog.  As expected from the comparisons of the previous section,
the U and $u$ filters are the most affected by the different metal
mixtures.  The BC for the reference mixture is systematically higher
by an amount that depends on the evolutionary phase. In both U and $u$
bands $\Delta$BC decreases moving from the low main MS towards the TO
(where it is practically negligible) then increases again along the
RGB, reaching a local maximum around $\teff=4476$~K and \glog=1.2,
before decreasing slowly towards the tip of the RGB. Here, the largest
values of $\Delta$BC are of the order of 0.2~mag in $u$ and 0.1~mag in
U.  The effect on the other filters is much smaller: the largest
values of $\Delta$BC are attained along the bright RGB and are of the
order of 0.02-0.04~mag.  Figure~\ref{BC2} displays the same comparison
for the case of the CNONa1 mixture, the
mixture with enhanced C+N+O. The behaviour of $\Delta$BC is
qualitatively identical to the previous case, but the maximum values
of $\Delta$BC are increased. They are equal to $\sim$0.30-0.35~mag in
$u$, $\sim$0.2~mag in U, ~$\sim$0.1~mag in $v$ and $\sim$0.04~mag in
the remaining filters.  The final comparison among BCs is displayed in
Fig.~\ref{BC3}, that shows the differences between the CNONa1 mixture
with $Y=0.246$, and the same metal mixture with a much higher $Y=0.40$.
The values of $\Delta$BC are reduced compared to the case of varying
the metal mixture. $\Delta$BC is essentially zero in all filters from
the low MS to the lower RGB, then becomes more and more negative while
climbing the RGB. A higher helium abundance appears to increase
the value of BC at 
low $\teff$ and low gravities. The largest differences are obtained
for the U and $u$ bands, but are of the order of only 0.05~mag at
most.  These variations are small with respect to the effect of the
metal mixture in U and $u$, but much more comparable for the other
filters.  Overall, the results of Fig.~\ref{BC3} are broadly
consistent with the conclusions by \citet{gcb07}, who studied the
effect of enhanced He (for a scaled solar metal mixture) on the BCs
for the UBVRIJHK filters.

Finally, we have also investigated whether the values of $\Delta$BC
are affected by the choice of microturbulence, using the additional
'test' calculations listed in Table~\ref{parameters}.  Microturbulence
affects line absorption by reducing line saturation: the net effect is
that saturated lines absorb more flux when microturbulence is higher,
while unsaturated lines do not change. Since typical microturbulence
values vary with the stellar evolutionary phases, one might need to
take it into account in estimating $\Delta$BC.  In practice, we find
that such effect is negligible: changing the value of $\mathrm
V_{turb}$ from 2 km/s to 0.5 km/s changes the values of $\Delta$BC by
at most 0.01~mag.

Following the method outlined in Sect.~\ref{methods} we have then
performed the first fully consistent theoretical study of the effect of the
CNONa abundance anticorrelations on the MS-TO-RGB CMDs of Galactic
GCs.  In particular, {\sl we are now in the position to account for the
CNONa pattern both in interior models and bolometric corrections.}

\begin{figure}
\centering \includegraphics[width=8.6cm]{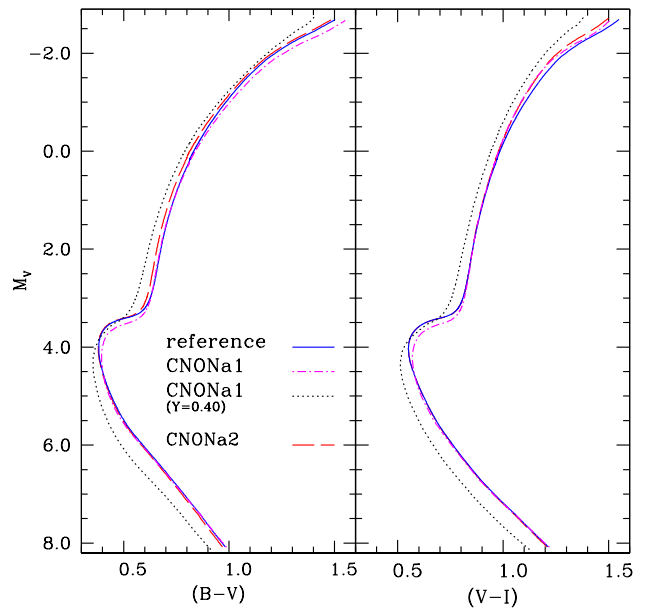}
\caption{{\sl Left:} The $M_{\rm V}$-(B$-$V) CMD for the same
  isochrones as in Fig.~\ref{HRiso}. Different line-styles correspond to
  different element mixtures as labelled.  {\sl Right:} As on the left,
  but for the $M_{\rm V}$-(V$-$I) CMD.  }
\label{BVI}
\end{figure}

\begin{figure}
\centering
\includegraphics[width=8.6cm]{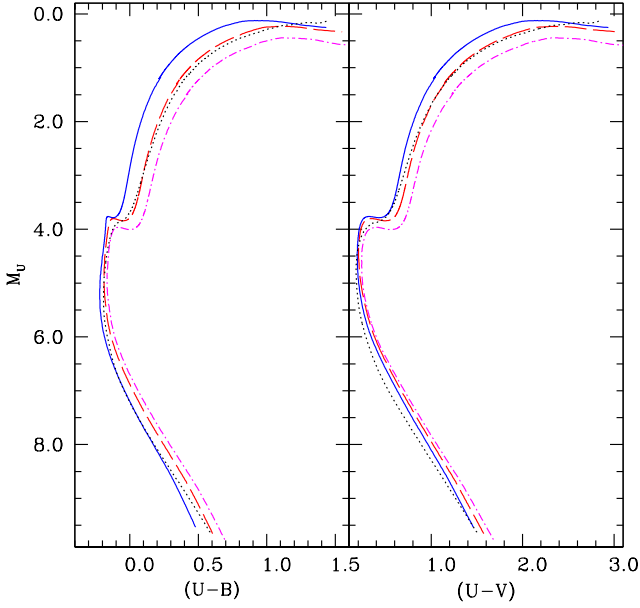}
\caption{As in Fig.~\ref{BVI} but for the  $M_{\rm U}$-(U$-$B) and 
$M_{\rm U}$-(U$-$V) CMDs.}
\label{UBV}
\end{figure}

\begin{figure}
\centering
\includegraphics[width=8.6cm]{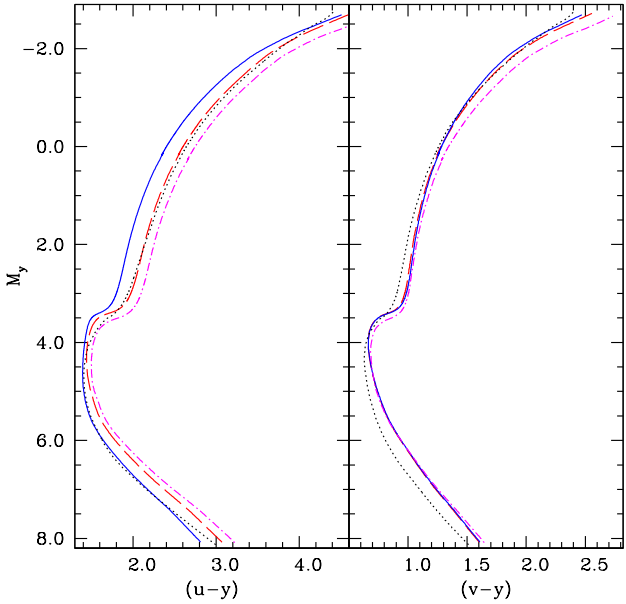}
\caption{As in Fig.~\ref{BVI} but for the  $M_y$-($v-y$) and 
$M_y$-($u-y$) CMDs.}
\label{uvy}
\end{figure}

\begin{figure}
\centering
\includegraphics[width=8.6cm]{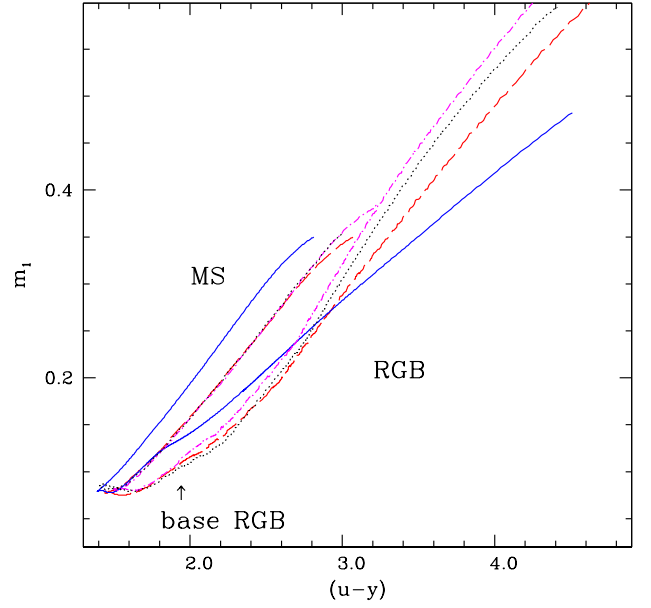}
\caption{As in Fig.~\ref{BVI} but for the  $m_1$-($u-y$) CMD.}
\label{m1}
\end{figure}

\begin{figure}
\centering
\includegraphics[width=8.6cm]{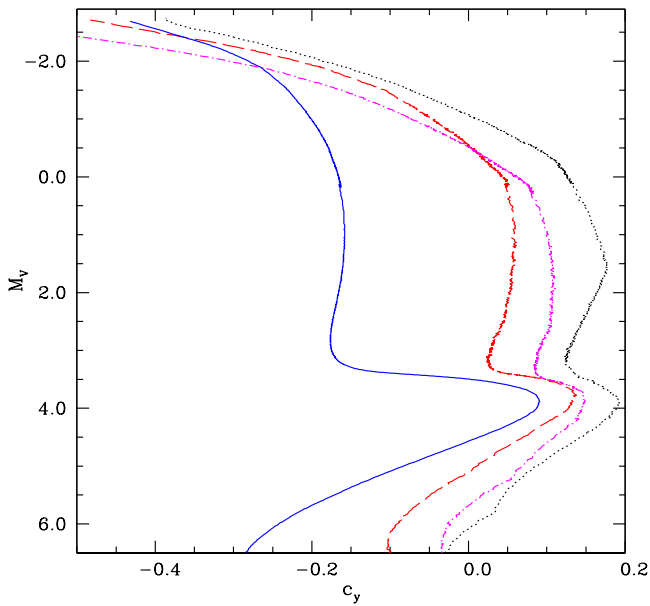}
\caption{As in Fig.~\ref{BVI} but for the  $M_{\rm V}$-($c_y$) CMD.}
\label{cy}
\end{figure}

Figures~\ref{BVI} to \ref{cy} display several CMDs in 
Johnson-Cousins as well as in Str\"omgren filters, for the isochrones of
Fig.~\ref{HRiso}. Given that the CNONa pattern we considered for the
second generation stars is characterised by extreme values for the
anticorrelation observed in Galactic GCs, the range of colours spanned
by our isochrones should give a rough idea -- considering also that
the extension of the abundance anticorrelations varies from cluster to
cluster \citep[e.g.][]{Carretta:10a} -- of the maximum colour spread
to be expected in the CMD of a generic Galactic GC.
 
The commonly used BVI diagrams are shown in Fig.~\ref{BVI}. The
behaviour of the isochrones in these CMDs mirrors closely the HRD of
Fig.~\ref{HRiso}, the reason being that in these filters the BCs are
hardly affected by the change in the metal mixture and initial He
content. As long as the sum of C+N+O is constant, second generation stars
are expected to overlap with first generation objects.  In case of
enhanced C+N+O, only the TO and SGB regions are affected, whereas MS and
RGB remain unchanged. An increase of $Y$ shifts the MS (and to a
lesser degree the RGB) towards bluer colours because of hotter $\teff$ in the evolutionary models.

The situation is -- not unexpectedly -- different when considering the
UBV CMDs displayed in Fig.~\ref{UBV}, given that U is the filter most
affected by the change of the metal mixture due to the emergence of strong molecular absorption in the atmosphere. In both diagrams the four
isochrones are well separated along the various branches. Overall, our
isochrone representative of second generation stars with enhanced C+N+O
is the reddest (dash-dotted line).  Notice how second generation stars
born from a metal mixture with constant C+N+O (long-dashed line)
follow a distinct sequence from the first generation population (solid
line), in spite of the fact that the HRDs of the 
underlying isochrones are identical. This is entirely due to the
effect of the metal mixture on the BCs.

The largest differences among the various isochrones appear along the
RGB, where the mixture with anticorrelations causes redder (U$-$B) and
(U$-$V) colours, at fixed $M_{\rm U}$. For example, at $M_{\rm U}$=2.0
the RGBs representative of second generation stars are redder by up to
$\sim$0.2~mag in (U-B) and $\sim$0.3~mag in (U$-$V), depending on the
metal mixture considered. We will compare this result with
observational evidence in the discussion section. 

It is also important to notice that an increase of $Y$ up to 0.40 -- a
very extreme He-enhancement -- shifts the isochrones of second
generation objects, which tend to be redder than the reference
isochrone due to the CNONa-variations, bluewards and thus closer again
to the reference isochrone. This is an effect of the stellar interior
models only, and may potentially even produce a bluer MS -- depending
on whether the anticorrelations are accompanied by a C+N+O enhancement
or whether the C+N+O sum stays unchanged -- especially in the
U-(U$-$V) diagrams. The effect can best be seen by comparing the
dot-dashed and dotted line in Fig.~\ref{UBV}.

As for the Str\"omgren filters, the $M_y$-($u-y$) and $M_y$-($v-y$) 
CMDs are displayed in Fig.~\ref{uvy}. The CMD with
($v-y$) colours show a behaviour very similar to the case of
the V-(B$-$V) plane, except for the RGB.  When 
the initial $Y$ is kept constant, the bright RGB for the C+N+O
enhanced mixture (dash-dotted line) is separated from the reference
isochrone (solid) -- at 
$M_y=-2.0$ the difference in ($v-y$) is $\sim$0.2~mag -- whereas the
mixture with anticorrelations at constant C+N+O (long-dashed) does not
modify the 
colour of the RGB.  This difference in the behaviour of the RGB ($v-y$)
colour for the two different mixtures with CNONa anticorrelations is
due essentially to the different BCs, rather than to the underlying
isochrones.

Finally, the $M_y$-($u-y$) CMD is similar qualitatively to the case of
the U-(U$-$B) diagram, with all different sequences well separated in
the CMD when the initial He is kept constant.

Figure~\ref{m1} displays the $m_1$-($u-y$) diagram, where the measured
$m_1=(v-b)-(b-y)$ in RGB stars is often used as an estimator of the
cluster metallicity \citep[see, e.g.,][and references
  therein]{calamida07}.  Our comparison shows that the presence of
CNONa anticorrelations produces a spread in the location of GC stars
in this diagram. Focusing our attention on RGB stars only (the right
part of the curves), the four
different isochrones in Fig.~\ref{m1} display quite different RGBs, which
intersect at ($u-y$) around 2.6-2.8~mag.  It is very interesting to
notice that even in case of a constant C+N+O sum, [Fe/H], $Y$ and total
metallicity $Z$ (the CNONa2 case; long-dashed line), the RGB
representative of second generation stars is 
different from the reference RGB, a clear effect of the difference in
BCs.

Finally, we consider the behaviour of the index $c_y$=$c_1-$($b-y$),
where $c_1=(u-v)-(v-b)$ is found empirically to be sensitive to the N
abundance \citep[see, e.g.,][]{yong08}. The $c_y$ index represents
well $c_1$, but removes much of the temperature sensitivity of this
index. As a result, empirical V-$c_y$ diagrams of Galactic GCs display
an almost vertical RGB at luminosities lower than the RGB bump
\citep{yong08}.  Figure~\ref{cy} shows how this empirical behaviour is
reproduced by the theoretical isochrones.  The theoretical RGBs of the
isochrones with our four selected chemical compositions are
approximately vertical below the bump luminosity, and are very well
separated in $c_y$ colour.  All isochrones representative of second
generation stars have redder $c_y$ colours, and the isochrone with the
largest N abundance (at fixed $Y$ -- the dot-dashed isochrone for the
CNONa1 mixture) is the reddest, in agreement with empirical results
\citep{lind10, yong08}. Notice how in this diagram an increase of $Y$
tends to move the RGBs of isochrones with CNONa anticorrelations
further away from the reference isochrone (dotted line), despite the
fact that the CNONa1 Y=0.4 mixture has a lower N abundance than the
CNONa1 Y=0.246 mixture.

\section{Discussion}

Using ATLAS~12 model atmospheres and SYNTHE spectrosyntheses we
calculated synthetic spectra for typical mixtures found in the various
subpopulations of globular clusters. From the theoretical spectra we
have determined bolometric corrections for standard Johnson-Cousins
and Str\"omgren filters, and finally predicted colours.

As a result of this work, we now have in hand complete and
self-consistent theoretical predictions of the effect of abundance
variations on observed CMDs, as they are found or discussed for the
different globular cluster stellar generations. Globally, element
abundance variations affect mainly the part of the spectra short of
about 400~nm due to changes in molecular bands. Therefore, colour
changes are largest in the blue filters and their detection is easiest
in the blue, independent of using broadband or narrow filters.
Enhanced helium abundance is affecting only the interior structure of
stars, leading to changes in luminosity and effective temperature, but
is irrelevant for the atmospheric structure at fixed $\log
\mathrm{g}$ and $T_\mathrm{eff}$, even for $Y$ as high as $0.400$.  

In this study, we did not investigate what the effect of
CNO and He abundance would be in cluster with different [Fe/H]. While
we don't expect the general behavior to change, the quantitative result is 
bound to be sensitive to the overall metal content of the cluster.

Using the combination of the previously computed isochrones and the
new bolometric corrections for the different GC subpopulations, we
summarize here the effects on the various possible colours in both
Johnson-Cousins and Str\"omgren filter systems:

\begin{enumerate} 
\item {\sl $BVI$-diagrams}: 
\begin{itemize} 
\item a splitting\footnote{In real globular clusters the splitting may
  in fact appear as a spread, depending on the nature of the element
  abundance variations} of sequences along the MS up
  to the TO, and to a lesser degree of the RGB can only be achieved --
  within the element variations discussed in this paper -- by varying
  the helium content $Y$. The CNONa anticorrelations influence neither 
  the stellar models nor the spectrum sufficiently when the C+N+O-abundance  
  is unchanged (mixture CNONa2 -- Fig.~\ref{BVI});
\item on the other hand, a variation of the C+N+O-abundance (mixture
  CNONa1) leads to a split of the SGB; this is entirely an effect of
  the stellar models. 
\end{itemize}
\item {\sl $UBV$- and  $uy$-diagrams}:
anticorrelations in CNONa abundances as well as $Y$-differences may
lead to multiple sequences from the MS to the RGB, where the effect
tends to be larger, and may reach 0.2--0.3~mag. This multiplicity is
independent of the sum of C+N+O (Fig.~\ref{UBV}; left panel). The
individual element variations are decisive. Helium enhancement,
however, works in the opposite direction than CNONa anticorrelations
(Fig.~\ref{uvy}; left panel).
\item {\sl $vy$-diagrams}: 
\begin{itemize} 
\item as in the case of the $BVI$-colours, a splitting of the MS up to
  the TO can be achieved only by a variation in $Y$;
\item similarly, after the TO, a split of the SGB is the result of a
  change in C+N+O;
\item additionally, a split along the RGB may result both from helium
 and from C+N+O variations; this is different from the $BVI$-case
(Fig.~\ref{uvy}; right panel).
\end{itemize}
\item {\sl $m_1$uy}-diagrams: 
\begin{itemize} 
\item CNONa anticorrelations lead to splits along the MS;
\item along the SG and RGB the same anticorrelations, but also
  helium variations lead to colour differences;
\item the sign of the colour change is different for the
  lower and upper part of the RGB (Fig.~\ref{m1}).
\end{itemize}
\item {\sl $c_y$V}-diagrams:
here, all parts of a CMD show the influence of both element
anticorrelations and of helium variations, and a strong separation can
be seen (Fig.~\ref{cy}).
\end{enumerate}

Our calculations also enable us to briefly address the issue
raised by the photometric observations by \citet{lee09}, who find
broad distributions of the $hk$ index -- defined as
$hk$=($Ca-b$)-($b-y$) -- in RGB sequences of several
Galactic GCs; they determined a broadening of the order of
$\sim$0.2~mag in $hk$ at fixed $V$ magnitude and attribute this to a
spread in calcium abundance, given that the $hk$ index is
sensitive to the ionized calcium H and K lines.  On the other hand,
direct spectroscopic measurements of calcium abundance by
\citet{carretta10} do not find any significant spread in individual
clusters.  We have tested with our spectra whether the CNONa abundance
anticorrelations affect the $hk$ index at our chosen RGB key
points. For this purpose we 
have calculated differences in the $hk$ index among our adopted
chemical compositions employing the appropriate profile from the $Ca$
filter taken from the Asiago photometric database \citep{fiorucci03}.
We find an effect of at most only $\sim$0.04~mag in the $hk$ colour at
fixed $M_{\rm V}$, second generation stars being systematically
bluer. Therefore, CNONa anticorrelations cannot fully explain 
$hk$-variations of 0.2~mag.

In comparison with observations we find that our M$_V-c_y$-diagram
closely follows empirical results by \citet{yong08} and
\citet{lind10}, which adds support to our theoretical models. We
thereby also confirm that this index is well suited to 
investigate photometrically variations of nitrogen abundance along the
lower RGB.

Also the U-(U$-$B) CMD of Fig.~\ref{UBV} interestingly agrees 
with the results by \citet{marinom4:08} about \object{M4} RGB
stars.  These authors 
have found that in the U-(U$-$B) CMD, objects with high Na --
represented by our isochrones for the two compositions with CNONa
anticorrelations -- are distributed to the red of stars with low Na --
represented by our isochrone with the reference chemical
composition.

A further interesting case is given by \object{NGC~6752}. In Sect.~1 we already
mentioned that \citet{milone:10} report about a MS splitting in visual
HST filters, and an RGB split in a UB-CMD. The SGB appears to be
single in the visual. \citet{yong08} found a variation in nitrogen
abundance of up to 1.95~dex, correlating with the $c_1$
index. Recently, \citet{kravtsov11} claimed a splitting of the SGB in
the U vs.\ B-I diagram, and a broadening of the RGB in the U vs.\ U-B
plane. The single SGB in B-I is indicative of a constant C+N+O, the
MS splitting of He-enhancement in the second generation stars, and the
broadening of the RGB in blue filters agrees with both. In
conclusions, if these findings are confirmed, they would indicate a
subpopulation of CNONa2 type, but with enhanced helium. 

Finally, we comment on the case of \object{NGC~1851} that has been observed in
different bands. The $y$-($v$-$y$), $y$-($b$-$y$), and $m_1$-($u$-$y$)
CMDs by \citet{calamida07} show a spread, if not a bimodality, along
the RGB.  The HST $m_\mathrm{F606W}$-($m_\mathrm{F606W}-m_\mathrm{F814W}$)
diagram by \citet{milone1851:08}, approximately equivalent to our
M$_V$ vs.\ V-I CMD, displays a double SGB, but no other splitting.  
In the UBVI observations of \citet{han09}, finally,
no RGB split is observed in the M$_V$ vs.\ V-I CMD, while
it is quite evident in the M$_U$ vs.\ (U-I) one \citep[see][Fig. 1]{han09}. It has to be noted that 
\citet{han09} could not detect any SGB split in the M$_V$ vs.\ V-I CMD, probably due to the 
lower quality of their ground-based photometry compared to the HST-based one in \citet{milone1851:08}.
On the basis of our results, only a second generation population with
enhanced C+N+O and `normal' He can satisfy all these constraints.
In fact, a second generation with constant C+N+O will not
produce a spread along the RGB in the $vy$ CMD, and a He enhancement
would reveal itself along the MS of the
$m_\mathrm{F606W}$-($m_\mathrm{F606W}-m_\mathrm{F814W}$) CMD.  Also
the synthetic HB simulations by \citet{salaris1851} seem to exclude a
He variation among the cluster subpopulations.

In summary, we are now able to investigate in detail clusters with known or
suspected abundance variations and several subpopulations, using 
information from their CMDs, on the basis of fully self-consistent interior and atmosphere models. As a
first application, in a following paper 
we will analyze in detail the cluster \object{NGC~1851} with more specific calculations. 

\begin{acknowledgements}
L.S.\ warmly acknowledges F.~Castelli for her prompt help with a
number of ATLAS12-related problems, as well as A. F. Marino for a number of comments and useful discussions of the paper findings. M.S.\ is grateful to the Max
Planck Institute for Astrophysics for their hospitality and
support. A.W.\ received financial support from the {\sl Excellence
  cluster ``Origin and Structure of the Universe''}
(Garching). S.C.\ acknowledges the financial support of INAF through
the PRIN INAF 2009 (P.I.: R. Gratton) .
\end{acknowledgements}
\bibliographystyle{aa}
\bibliography{refColours}

\Online
\clearpage
\onecolumn

{\tiny
\begin{longtable}{rll r@{.}l l r@{.}l r@{.}l l r@{.}l r@{.}l lr@{.}l r@{.}l}
\caption{Abundances for all the elements in the considered mixtures.\label{fullmixtures_full}}  \\
\hline    
 & & \multicolumn{3}{l}{REFERENCE $Y$=0.246}&\multicolumn{5}{l}{CNONa1 $Y$=0.246}&\multicolumn{5}{l}{CNONa1 $Y$=0.400}&\multicolumn{5}{l}{CNONa2 $Y$=0.246} \\ 
\hline
N   & El.& Mass frac. &  \multicolumn{2}{l}{[El]$^{a}$}&   Mass frac. &
\multicolumn{2}{l}{[El]}& \multicolumn{2}{l}{[El]-[El]$_{ref}^{b}$} & Mass frac. &
\multicolumn{2}{l}{[El]}  & \multicolumn{2}{l}{[El]-[El]$_{ref}$} & Mass
frac. & \multicolumn{2}{l}{[El]}  & \multicolumn{2}{l}{[El]-[El]$_{ref}$} \\ 
\hline

 1 & H  & 7.5300e-01 &  12&000 & 7.5217e-01 &  12&000 &  0&0000 & 5.9854e-01 &  12&000 &  0&0000 & 7.5300e-01 &  12&000 &  0&0000 \\
 2 & He & 2.4600e-01 &  10&915 & 2.4600e-01 &  10&916 &  0&0005 & 4.0000e-01 &  11&226 &  0&3108 & 2.4600e-01 &  10&915 &  0&0000 \\
 3 & Li & 1.8376e-12 &  -0&451 & 3.3729e-12 &  -0&186 &  0&2642 & 2.6840e-12 &  -0&186 &  0&2642 & 1.8376e-12 &  -0&451 &  0&0000 \\
 4 & Be & 4.7607e-12 &  -0&151 & 8.7380e-12 &   0&114 &  0&2642 & 6.9534e-12 &   0&114 &  0&2642 & 4.7607e-12 &  -0&151 &  0&0000 \\
 5 & B  & 8.0668e-11 &   0&999 & 1.4806e-10 &   1&264 &  0&2642 & 1.1782e-10 &   1&264 &  0&2642 & 8.0668e-11 &   0&999 &  0&0000 \\
 6 & C  & 7.6386e-05 &   6&930 & 1.9172e-05 &   6&330 & -0&5999 & 1.5256e-05 &   6&330 & -0&5999 & 1.9184e-05 &   6&330 & -0&6001 \\
 7 & N  & 2.3430e-05 &   6&350 & 1.4768e-03 &   8&150 &  1&8000 & 1.1752e-03 &   8&150 &  1&8000 & 6.4150e-04 &   7&787 &  1&4374 \\
 8 & O  & 6.7226e-04 &   7&750 & 1.0642e-04 &   6&950 & -0&8000 & 8.4687e-05 &   6&950 & -0&8000 & 1.0671e-04 &   6&951 & -0&7993 \\
 9 & F  & 1.4506e-08 &   3&009 & 2.6626e-08 &   3&274 &  0&2642 & 2.1188e-08 &   3&274 &  0&2642 & 1.4506e-08 &   3&009 &  0&0000 \\
10 & Ne & 8.4796e-05 &   6&750 & 8.4706e-05 &   6&750 &  0&0000 & 6.7405e-05 &   6&750 &  0&0000 & 8.4796e-05 &   6&750 &  0&0000 \\
11 & Na & 8.8125e-07 &   4&710 & 5.5531e-06 &   5&510 &  0&7999 & 4.4189e-06 &   5&510 &  0&7999 & 5.5602e-06 &   5&510 &  0&8000 \\
12 & Mg & 4.1603e-05 &   6&360 & 4.1558e-05 &   6&360 &  0&0000 & 3.3070e-05 &   6&360 &  0&0000 & 4.1603e-05 &   6&360 &  0&0000 \\
13 & Al & 1.4268e-06 &   4&850 & 1.4249e-06 &   4&850 & -0&0001 & 1.1339e-06 &   4&850 & -0&0001 & 1.4268e-06 &   4&850 &  0&0000 \\
14 & Si & 3.5638e-05 &   6&230 & 3.5600e-05 &   6&230 &  0&0000 & 2.8329e-05 &   6&230 &  0&0000 & 3.5638e-05 &   6&230 &  0&0000 \\
15 & P  & 1.5687e-07 &   3&831 & 1.5588e-07 &   3&829 & -0&0023 & 1.2404e-07 &   3&829 & -0&0023 & 1.5687e-07 &   3&831 &  0&0000 \\
16 & S  & 1.9925e-05 &   5&920 & 1.9903e-05 &   5&920 &  0&0000 & 1.5838e-05 &   5&920 &  0&0000 & 1.9925e-05 &   5&920 &  0&0000 \\
17 & Cl & 2.0083e-07 &   3&880 & 1.9990e-07 &   3&878 & -0&0015 & 1.5907e-07 &   3&878 & -0&0015 & 2.0083e-07 &   3&880 &  0&0000 \\
18 & Ar & 2.3710e-06 &   4&900 & 2.3694e-06 &   4&900 &  0&0002 & 1.8855e-06 &   4&900 &  0&0002 & 2.3710e-06 &   4&900 &  0&0000 \\
19 & K  & 9.1921e-08 &   3&498 & 9.1695e-08 &   3&497 & -0&0006 & 7.2967e-08 &   3&497 & -0&0006 & 9.1921e-08 &   3&498 &  0&0000 \\
20 & Ca & 5.2045e-06 &   5&240 & 5.2009e-06 &   5&240 &  0&0002 & 4.1387e-06 &   5&240 &  0&0002 & 5.2045e-06 &   5&240 &  0&0000 \\
21 & Sc & 1.3984e-09 &   1&619 & 2.5667e-09 &   1&884 &  0&2642 & 2.0425e-09 &   1&884 &  0&2642 & 1.3984e-09 &   1&619 &  0&0000 \\
22 & Ti & 3.8667e-07 &   4&034 & 3.8512e-07 &   4&033 & -0&0013 & 3.0646e-07 &   4&033 & -0&0013 & 3.8667e-07 &   4&034 &  0&0000 \\
23 & V  & 1.0713e-08 &   2&449 & 1.9663e-08 &   2&714 &  0&2642 & 1.5647e-08 &   2&714 &  0&2642 & 1.0713e-08 &   2&449 &  0&0000 \\
24 & Cr & 4.4262e-07 &   4&057 & 4.4197e-07 &   4&057 & -0&0002 & 3.5170e-07 &   4&057 & -0&0002 & 4.4262e-07 &   4&057 &  0&0000 \\
25 & Mn & 2.4179e-07 &   3&770 & 2.4208e-07 &   3&771 &  0&0010 & 1.9263e-07 &   3&771 &  0&0010 & 2.4179e-07 &   3&770 &  0&0000 \\
26 & Fe & 3.1648e-05 &   5&880 & 3.1613e-05 &   5&880 &  0&0000 & 2.5156e-05 &   5&880 &  0&0000 & 3.1648e-05 &   5&880 &  0&0000 \\
27 & Co & 1.0309e-07 &   3&369 & 1.8921e-07 &   3&634 &  0&2642 & 1.5057e-07 &   3&634 &  0&2642 & 1.0309e-07 &   3&369 &  0&0000 \\
28 & Ni & 2.0542e-06 &   4&671 & 2.0503e-06 &   4&670 & -0&0003 & 1.6315e-06 &   4&670 & -0&0003 & 2.0542e-06 &   4&671 &  0&0000 \\
29 & Cu & 2.1673e-08 &   2&659 & 3.9781e-08 &   2&924 &  0&2642 & 3.1656e-08 &   2&924 &  0&2642 & 2.1673e-08 &   2&659 &  0&0000 \\
30 & Zn & 5.4737e-08 &   3&049 & 1.0047e-07 &   3&314 &  0&2642 & 7.9948e-08 &   3&314 &  0&2642 & 5.4737e-08 &   3&049 &  0&0000 \\
31 & Ga & 1.1123e-09 &   1&329 & 2.0416e-09 &   1&594 &  0&2642 & 1.6246e-09 &   1&594 &  0&2642 & 1.1123e-09 &   1&329 &  0&0000 \\
32 & Ge & 3.9266e-09 &   1&859 & 7.2071e-09 &   2&124 &  0&2642 & 5.7351e-09 &   2&124 &  0&2642 & 3.9266e-09 &   1&859 &  0&0000 \\
33 & As & 3.6935e-10 &   0&819 & 6.7794e-10 &   1&084 &  0&2642 & 5.3948e-10 &   1&084 &  0&2642 & 3.6935e-10 &   0&819 &  0&0000 \\
34 & Se & 4.2682e-09 &   1&859 & 7.8341e-09 &   2&124 &  0&2642 & 6.2341e-09 &   2&124 &  0&2642 & 4.2682e-09 &   1&859 &  0&0000 \\
35 & Br & 7.1681e-10 &   1&079 & 1.3157e-09 &   1&344 &  0&2642 & 1.0470e-09 &   1&344 &  0&2642 & 7.1681e-10 &   1&079 &  0&0000 \\
36 & Kr & 3.5981e-09 &   1&759 & 6.6042e-09 &   2&024 &  0&2642 & 5.2553e-09 &   2&024 &  0&2642 & 3.5981e-09 &   1&759 &  0&0000 \\
37 & Rb & 7.1555e-10 &   1&049 & 1.3134e-09 &   1&314 &  0&2642 & 1.0451e-09 &   1&314 &  0&2642 & 7.1555e-10 &   1&049 &  0&0000 \\
38 & Sr & 1.7197e-09 &   1&419 & 3.1564e-09 &   1&684 &  0&2642 & 2.5117e-09 &   1&684 &  0&2642 & 1.7197e-09 &   1&419 &  0&0000 \\
39 & Y  & 3.2491e-10 &   0&689 & 5.9637e-10 &   0&954 &  0&2642 & 4.7456e-10 &   0&954 &  0&2642 & 3.2491e-10 &   0&689 &  0&0000 \\
40 & Zr & 7.6374e-10 &   1&049 & 1.4018e-09 &   1&314 &  0&2642 & 1.1155e-09 &   1&314 &  0&2642 & 7.6374e-10 &   1&049 &  0&0000 \\
41 & Nb & 5.1390e-11 &  -0&131 & 9.4326e-11 &   0&134 &  0&2642 & 7.5060e-11 &   0&134 &  0&2642 & 5.1390e-11 &  -0&131 &  0&0000 \\
42 & Mo & 1.6785e-10 &   0&369 & 3.0809e-10 &   0&634 &  0&2642 & 2.4516e-10 &   0&634 &  0&2642 & 1.6785e-10 &   0&369 &  0&0000 \\
43 & Tc & 2.2598e-20 &  -9&511 & 4.1477e-20 &  -9&246 &  0&2642 & 3.3006e-20 &  -9&246 &  0&2642 & 2.2598e-20 &  -9&511 &  0&0000 \\
44 & Ru & 1.4705e-10 &   0&289 & 2.6990e-10 &   0&554 &  0&2642 & 2.1478e-10 &   0&554 &  0&2642 & 1.4705e-10 &   0&289 &  0&0000 \\
45 & Rh & 2.8528e-11 &  -0&431 & 5.2363e-11 &  -0&166 &  0&2642 & 4.1668e-11 &  -0&166 &  0&2642 & 2.8528e-11 &  -0&431 &  0&0000 \\
46 & Pd & 1.0961e-10 &   0&139 & 2.0119e-10 &   0&404 &  0&2642 & 1.6010e-10 &   0&404 &  0&2642 & 1.0961e-10 &   0&139 &  0&0000 \\
47 & Ag & 1.9757e-11 &  -0&611 & 3.6264e-11 &  -0&346 &  0&2642 & 2.8857e-11 &  -0&346 &  0&2642 & 1.9757e-11 &  -0&611 &  0&0000 \\
48 & Cd & 1.3920e-10 &   0&219 & 2.5550e-10 &   0&484 &  0&2642 & 2.0332e-10 &   0&484 &  0&2642 & 1.3920e-10 &   0&219 &  0&0000 \\
49 & In & 1.1037e-10 &   0&109 & 2.0258e-10 &   0&374 &  0&2642 & 1.6120e-10 &   0&374 &  0&2642 & 1.1037e-10 &   0&109 &  0&0000 \\
50 & Sn & 2.4965e-10 &   0&449 & 4.5822e-10 &   0&714 &  0&2642 & 3.6463e-10 &   0&714 &  0&2642 & 2.4965e-10 &   0&449 &  0&0000 \\
51 & Sb & 2.5606e-11 &  -0&551 & 4.6999e-11 &  -0&286 &  0&2642 & 3.7400e-11 &  -0&286 &  0&2642 & 2.5606e-11 &  -0&551 &  0&0000 \\
52 & Te & 4.6632e-10 &   0&689 & 8.5592e-10 &   0&954 &  0&2642 & 6.8111e-10 &   0&954 &  0&2642 & 4.6632e-10 &   0&689 &  0&0000 \\
53 & I  & 8.6360e-11 &  -0&041 & 1.5851e-10 &   0&224 &  0&2642 & 1.2614e-10 &   0&224 &  0&2642 & 8.6360e-11 &  -0&041 &  0&0000 \\
54 & Xe & 4.0839e-10 &   0&619 & 7.4959e-10 &   0&884 &  0&2642 & 5.9649e-10 &   0&884 &  0&2642 & 4.0839e-10 &   0&619 &  0&0000 \\
55 & Cs & 3.7703e-11 &  -0&421 & 6.9203e-11 &  -0&156 &  0&2642 & 5.5069e-11 &  -0&156 &  0&2642 & 3.7703e-11 &  -0&421 &  0&0000 \\
56 & Ba & 3.8958e-10 &   0&579 & 7.1506e-10 &   0&844 &  0&2642 & 5.6901e-10 &   0&844 &  0&2642 & 3.8958e-10 &   0&579 &  0&0000 \\
57 & La & 4.3207e-11 &  -0&381 & 7.9306e-11 &  -0&116 &  0&2642 & 6.3108e-11 &  -0&116 &  0&2642 & 4.3207e-11 &  -0&381 &  0&0000 \\
58 & Ce & 1.1203e-10 &   0&029 & 2.0562e-10 &   0&294 &  0&2642 & 1.6363e-10 &   0&294 &  0&2642 & 1.1203e-10 &   0&029 &  0&0000 \\
59 & Pr & 1.5197e-11 &  -0&841 & 2.7895e-11 &  -0&576 &  0&2642 & 2.2197e-11 &  -0&576 &  0&2642 & 1.5197e-11 &  -0&841 &  0&0000 \\
60 & Nd & 9.5924e-11 &  -0&051 & 1.7607e-10 &   0&214 &  0&2642 & 1.4011e-10 &   0&214 &  0&2642 & 9.5924e-11 &  -0&051 &  0&0000 \\
61 & Pm & 3.3435e-20 &  -9&511 & 6.1369e-20 &  -9&246 &  0&2642 & 4.8835e-20 &  -9&246 &  0&2642 & 3.3435e-20 &  -9&511 &  0&0000 \\
62 & Sm & 3.2357e-11 &  -0&541 & 5.9391e-11 &  -0&276 &  0&2642 & 4.7260e-11 &  -0&276 &  0&2642 & 3.2357e-11 &  -0&541 &  0&0000 \\
63 & Eu & 1.0341e-11 &  -1&041 & 1.8981e-11 &  -0&776 &  0&2642 & 1.5104e-11 &  -0&776 &  0&2642 & 1.0341e-11 &  -1&041 &  0&0000 \\
64 & Gd & 4.3594e-11 &  -0&431 & 8.0016e-11 &  -0&166 &  0&2642 & 6.3673e-11 &  -0&166 &  0&2642 & 4.3594e-11 &  -0&431 &  0&0000 \\
65 & Tb & 7.4822e-12 &  -1&201 & 1.3733e-11 &  -0&936 &  0&2642 & 1.0928e-11 &  -0&936 &  0&2642 & 7.4822e-12 &  -1&201 &  0&0000 \\
66 & Dy & 4.7173e-11 &  -0&411 & 8.6584e-11 &  -0&146 &  0&2642 & 6.8900e-11 &  -0&146 &  0&2642 & 4.7173e-11 &  -0&411 &  0&0000 \\
67 & Ho & 6.3116e-12 &  -1&291 & 1.1585e-11 &  -1&026 &  0&2642 & 9.2186e-12 &  -1&026 &  0&2642 & 6.3116e-12 &  -1&291 &  0&0000 \\
68 & Er & 2.9938e-11 &  -0&621 & 5.4951e-11 &  -0&356 &  0&2642 & 4.3727e-11 &  -0&356 &  0&2642 & 2.9938e-11 &  -0&621 &  0&0000 \\
69 & Tm & 3.5527e-12 &  -1&551 & 6.5208e-12 &  -1&286 &  0&2642 & 5.1890e-12 &  -1&286 &  0&2642 & 3.5527e-12 &  -1&551 &  0&0000 \\
70 & Yb & 4.3754e-11 &  -0&471 & 8.0309e-11 &  -0&206 &  0&2642 & 6.3907e-11 &  -0&206 &  0&2642 & 4.3754e-11 &  -0&471 &  0&0000 \\
71 & Lu & 4.2247e-12 &  -1&491 & 7.7543e-12 &  -1&226 &  0&2642 & 6.1705e-12 &  -1&226 &  0&2642 & 4.2247e-12 &  -1&491 &  0&0000 \\
72 & Hf & 2.8474e-11 &  -0&671 & 5.2264e-11 &  -0&406 &  0&2642 & 4.1589e-11 &  -0&406 &  0&2642 & 2.8474e-11 &  -0&671 &  0&0000 \\
73 & Ta & 2.8209e-12 &  -1&681 & 5.1777e-12 &  -1&416 &  0&2642 & 4.1202e-12 &  -1&416 &  0&2642 & 2.8209e-12 &  -1&681 &  0&0000 \\
74 & W  & 4.9805e-11 &  -0&441 & 9.1416e-11 &  -0&176 &  0&2642 & 7.2745e-11 &  -0&176 &  0&2642 & 4.9805e-11 &  -0&441 &  0&0000 \\
75 & Re & 7.4616e-12 &  -1&271 & 1.3696e-11 &  -1&006 &  0&2642 & 1.0898e-11 &  -1&006 &  0&2642 & 7.4616e-12 &  -1&271 &  0&0000 \\
76 & Os & 1.1275e-10 &  -0&101 & 2.0695e-10 &   0&164 &  0&2642 & 1.6468e-10 &   0&164 &  0&2642 & 1.1275e-10 &  -0&101 &  0&0000 \\
77 & Ir & 9.0496e-11 &  -0&201 & 1.6610e-10 &   0&064 &  0&2642 & 1.3218e-10 &   0&064 &  0&2642 & 9.0496e-11 &  -0&201 &  0&0000 \\
78 & Pt & 2.5886e-10 &   0&249 & 4.7512e-10 &   0&514 &  0&2642 & 3.7808e-10 &   0&514 &  0&2642 & 2.5886e-10 &   0&249 &  0&0000 \\
79 & Au & 4.2387e-11 &  -0&541 & 7.7800e-11 &  -0&276 &  0&2642 & 6.1909e-11 &  -0&276 &  0&2642 & 4.2387e-11 &  -0&541 &  0&0000 \\
80 & Hg & 5.6904e-11 &  -0&421 & 1.0445e-10 &  -0&156 &  0&2642 & 8.3114e-11 &  -0&156 &  0&2642 & 5.6904e-11 &  -0&421 &  0&0000 \\
81 & Tl & 3.4141e-11 &  -0&651 & 6.2666e-11 &  -0&386 &  0&2642 & 4.9867e-11 &  -0&386 &  0&2642 & 3.4141e-11 &  -0&651 &  0&0000 \\
82 & Pb & 3.8835e-10 &   0&399 & 7.1281e-10 &   0&664 &  0&2642 & 5.6722e-10 &   0&664 &  0&2642 & 3.8835e-10 &   0&399 &  0&0000 \\
83 & Bi & 2.2539e-11 &  -0&841 & 4.1370e-11 &  -0&576 &  0&2642 & 3.2921e-11 &  -0&576 &  0&2642 & 2.2539e-11 &  -0&841 &  0&0000 \\
84 & Po & 4.8193e-20 &  -9&511 & 8.8457e-20 &  -9&246 &  0&2642 & 7.0390e-20 &  -9&246 &  0&2642 & 4.8193e-20 &  -9&511 &  0&0000 \\
85 & At & 4.8423e-20 &  -9&511 & 8.8880e-20 &  -9&246 &  0&2642 & 7.0727e-20 &  -9&246 &  0&2642 & 4.8423e-20 &  -9&511 &  0&0000 \\
86 & Rn & 5.1191e-20 &  -9&511 & 9.3959e-20 &  -9&246 &  0&2642 & 7.4768e-20 &  -9&246 &  0&2642 & 5.1191e-20 &  -9&511 &  0&0000 \\
87 & Fr & 5.1421e-20 &  -9&511 & 9.4382e-20 &  -9&246 &  0&2642 & 7.5105e-20 &  -9&246 &  0&2642 & 5.1421e-20 &  -9&511 &  0&0000 \\
88 & Ra & 5.2113e-20 &  -9&511 & 9.5652e-20 &  -9&246 &  0&2642 & 7.6115e-20 &  -9&246 &  0&2642 & 5.2113e-20 &  -9&511 &  0&0000 \\
89 & Ac & 5.2343e-20 &  -9&511 & 9.6075e-20 &  -9&246 &  0&2642 & 7.6452e-20 &  -9&246 &  0&2642 & 5.2343e-20 &  -9&511 &  0&0000 \\
90 & Th & 6.0034e-12 &  -1&461 & 1.1019e-11 &  -1&196 &  0&2642 & 8.7685e-12 &  -1&196 &  0&2642 & 6.0034e-12 &  -1&461 &  0&0000 \\
91 & Pa & 5.3274e-20 &  -9&511 & 9.7783e-20 &  -9&246 &  0&2642 & 7.7812e-20 &  -9&246 &  0&2642 & 5.3274e-20 &  -9&511 &  0&0000 \\
92 & U  & 1.5829e-12 &  -2&051 & 2.9055e-12 &  -1&786 &  0&2642 & 2.3120e-12 &  -1&786 &  0&2642 & 1.5829e-12 &  -2&051 &  0&0000 \\
93 & Np & 5.4649e-20 &  -9&511 & 1.0031e-19 &  -9&246 &  0&2642 & 7.9820e-20 &  -9&246 &  0&2642 & 5.4649e-20 &  -9&511 &  0&0000 \\
94 & Pu & 5.6263e-20 &  -9&511 & 1.0327e-19 &  -9&246 &  0&2642 & 8.2178e-20 &  -9&246 &  0&2642 & 5.6263e-20 &  -9&511 &  0&0000 \\
95 & Am & 5.6033e-20 &  -9&511 & 1.0285e-19 &  -9&246 &  0&2642 & 8.1841e-20 &  -9&246 &  0&2642 & 5.6033e-20 &  -9&511 &  0&0000 \\
96 & Cm & 5.6955e-20 &  -9&511 & 1.0454e-19 &  -9&246 &  0&2642 & 8.3188e-20 &  -9&246 &  0&2642 & 5.6955e-20 &  -9&511 &  0&0000 \\
97 & Bk & 5.6955e-20 &  -9&511 & 1.0454e-19 &  -9&246 &  0&2642 & 8.3188e-20 &  -9&246 &  0&2642 & 5.6955e-20 &  -9&511 &  0&0000 \\
98 & Cf & 5.7878e-20 &  -9&511 & 1.0623e-19 &  -9&246 &  0&2642 & 8.4535e-20 &  -9&246 &  0&2642 & 5.7878e-20 &  -9&511 &  0&0000 \\
99 & Es & 5.8108e-20 &  -9&511 & 1.0666e-19 &  -9&246 &  0&2642 & 8.4872e-20 &  -9&246 &  0&2642 & 5.8108e-20 &  -9&511 &  0&0000 \\
\hline    
   & $Z$   & \multicolumn{3}{l}{{\bf 9.9937e-04}}      &
\multicolumn{5}{l}{{\bf 1.8343e-03}}&               \multicolumn{5}{l}{
 {\bf 1.4597e-03}}               &   \multicolumn{5}{l}{{\bf 9.9937e-04}} \\ 
\hline
\multicolumn{14}{l}{\scriptsize $a$ [El]=$\log N(\mathrm{El})-\log N(\mathrm{H})+12$}\\
\multicolumn{14}{l}{\scriptsize $b$ [El] for this mixture minus [El] for the reference mixture.}\\
\end{longtable}
}

\end{document}